\documentstyle[12pt]{article}
\date{}
\begin{document}

\title{Mesonic Anapole Form Factors of the Nucleons}

\author{D.O. Riska}
\maketitle

\centerline{\it Department of Physics, POB 9, 00014 University of 
Helsinki, Finland}

\begin{abstract}
The chiral quark model posits pseudoscalar and vector meson 
couplings to constituent quarks. The parity violating meson-quark 
couplings
lead to anapole moments and form factors of the 
nucleons. These arise both as parity
violating meson loop fluctuations as well as exchange currents
and polarization currents that are induced by the parity
violating interaction between quarks. Because of cancellations 
between the different contributions the
magnitude of the calculated anapole moments is only of the order $\sim
10^{-8}$ and is
determined mainly by magnitude of the parity-violating
meson-nucleon coupling constants.
\end{abstract}
\newpage

\section{Introduction}

Experimental determination of the strangeness form factors of the
nucleons aims at the coupling of the $Z^0$ boson to the strange quarks
and measures the interference term between the $\gamma$ and $Z^0$
exchange interactions between electrons and protons in electron-nucleons
scattering \cite{SAMPLE1,SAMPLE2,HAPPEX}. If the nucleons have non-zero
anapole moments, these contribute an axial current component to the
electron-nucleon coupling, which has to be taken into account as a
radiative
correction in the extraction of the strangeness form factors from the
measured asymmetry \cite{MusolfH}.

Anapole moments are induced by parity violating meson fluctuations of
the nucleons, as e.g. $W^\pm$ fluctuations \cite{MusolfT} and
$\pi$-loops with one parity violating (PV) vertex \cite{Henley1}. For the
nucleons the anapole form factors may be defined as
$F_A^{p,n}(0)=F_A^S(q^2)\pm F_A^V(q^2)$, where $F_A^S(q^2)$ and
$F_A^V(q^2)$ are the isoscalar and isovector form factors in the current
matrix element
$$<p'|j_\mu^A(0)|p>=ie\bar u(p'){F_A^S(q^2)+\tau_3F_A^V(q^2)\over m_N^2}
[q^2\gamma_\mu\gamma_5-2im_Nq_\mu\gamma_5]u(p).\eqno(1.1)$$
Here $m_N$ is the nucleon mass, and $q$ is the momentum transfer to the
nucleon ($q=p'-p)$. The space favoring metric $q^2=\vec q^2-q_0^2$
will be employed throughout.

The anapole moment due to pionic fluctuations of the proton was
estimated in ref. \cite{Henley1} to be $-0.12f_\pi$, where $f_\pi$ is
the PV pion-nucleon coupling constant, the standard value of which is
$-4.5\cdot 10^{-7}$ \cite{DDH,Kisslinger}. An anapole moment this small
does not affect the empirical extraction of the strangeness magnetic
moment of the proton in PV electron scattering. A significant effect
from the anapole moment would require that it be larger by at least an
order of magnitude.

A calculation of the anapole moment of the proton based on the chiral
quark model is reported here. As the pseudoscalar and vector mesons
couple directly to constituent quarks in this model, the anapole moment
arises as a sum of contributions from
PV meson loop fluctuations of the constituent quarks
and PV meson exchange currents along with PV "polarization currents"
induced by the PV meson exchange interaction between quarks.
Current conservation
demands the presence of all of these combined. The
calculation is therefore reminiscent
of the calculation of nuclear anapole moments in refs.
\cite{MusolfT,Henley1}. This provides an alternate approach to the
baryonic loop calculations, which recently have been recast into the
form of chiral perturbation theory \cite{Maekawa,Holstein00}. 
The calculation based
on the chiral quark model allows a unified treatment of baryon structure
and the baryon spectrum based on the same Hamiltonian. It
takes all baryonic intermediate states into account, as the constituent
quarks lack excited states.
A major difference between the quark model and effective baryon
field theory approaches is that while the $SU(6)$ aspect of the
baryon wave function plays 
plays a major role in the former it plays none whatever
in the latter.
The magnetic moments of the baryons provide
an example of
a set of observables for which the 3-quark structure of the
baryon wave function plays the leading role, and meson
loops give but small contributions in the quark model.

In the chiral quark model the magnitude of the calculated anapole moment
is determined by the PV meson-nucleon coupling constants. By taking into
account both the pion and vector meson exchange contributions, the
value
for the anapole moment of the proton is found to be
$-0.9\cdot 10^{-8}$, with a very large uncertainty margin, that
mostly is set by the poorly known values of the PV meson-nucleon
coupling constants. The pion contributions alone give rise
to a positive value for the anapole moment ($+1.88\cdot 10^{-8}$), 
in agreement with results that are obtained in chiral perturbation 
theory
\cite{Maekawa}.  
With the "recommended" values for the PV meson-nucleon
coupling constants \cite{DDH} there are strong cancellations between the
pion and vector meson contributions to the anapole moment.
The pionic contribution to the anapole moment is found to be
somewhat smaller in the chiral quark model than in the pion loop
calculation at the baryon level \cite{Henley1} because of a
tendency to cancellation
between the pion loop and pion exchange current contributions. 

This paper falls into 7 sections. In sections 2 and 3 the pion loop and
pion exchange and polarization current contributions
respectively are calculated. In
sections 4 and 5 the corresponding vector meson loop contributions are
derived. In section 6 the $\rho\pi$ and $\omega\pi$ loop
contributions are calculated.
Section 7 contains a discussions of the results and their
implications.

\section{Pion loop contributions to the anapole form factors}

The parity violating (PV) and parity conserving (PC) couplings to
constituent quarks have the expressions:
$${\cal L}_{\pi qq}^{PV}=g_{\pi qq}^W\bar \psi(\vec \tau\times \vec
\phi)_3\psi,\eqno(2.1a)$$
$${\cal L}_{\pi qq}^{PC}=i{f_{\pi qq}\over m_\pi}\bar
\psi\gamma_5\gamma_\mu\partial_\mu\vec \phi\cdot \vec \tau\psi.
\eqno(2.1b)$$
Here $\psi$ represents the field of the constituent $u$ and $d$ quarks
and $\vec\phi$ the pion field. By means of standard quark model algebra
the pion-quark coupling constants may be expressed in terms of the
corresponding pion-nucleon coupling constants as
$$g_{\pi qq}^W=g_{\pi NN}^W=-{f_\pi\over \sqrt{2}},\eqno(2.2a)$$
$$f_{\pi qq}^{PC}={3\over 5}f_{\pi NN}\simeq 0.6.\eqno(2.2b)$$
Here $f_\pi$ is the PV $\pi N$ coupling in the notation of ref. 
\cite{Holstein} and $f_{\pi NN}$ is the pseudovector $\pi N$-coupling.

The PV pion fluctuations of the constituent quarks, which contribute to
the anapole moments of the constituent quarks are illustrated by the
Feynman diagrams in Fig. 1. For the evaluation of these loop amplitudes
the pion and quark current operators are e
$$j_\mu ^\pi=e(\partial_\mu\vec \phi\times \vec \phi)_3,\eqno(2.3a)$$
$$j_\mu ^q=ie\bar\psi({1\over 6}+{\tau_3\over 2})\gamma_\mu
\psi.\eqno(2.3b)$$
Here it has been assumed that the anomalous Pauli terms in the e.m.
current of the constituent quarks are unimportant \cite{bira,Gloz}.

Minimal substitution of the e.m. field $\vec A$ in the chiral coupling
(2.1b) generates a contact coupling term:
$${\cal L}_{\pi \gamma qq}=ie{f_{\pi qq}\over m_\pi}\bar\psi
\gamma_5\gamma_\mu
A_\mu(\vec \phi\times \vec\tau)_3\psi,\eqno(2.4)$$
which generates contact coupling diagrams in addition to the pion loop
diagrams in Fig. 1. For e.m. couplings, the amplitudes that are obtained
with the PV coupling (2.1b) and the contact coupling (2.4) are
equivalent to those obtained with the pseudoscalar coupling
$${\cal L}_{\pi qq}=ig_{\pi qq}\bar\psi \gamma_5\vec \phi\cdot \vec \tau
\psi,\eqno(2.5)$$
if $g_{\pi qq}=(2m_q/m_\pi) f_{\pi qq}$ ($m$ is the constituent
quark mass).

The contributions from the pion loop fluctuations in Figs. 1 to the e.m.
current of the $u[d]$ quarks may be expressed as
$$j_\mu=-[+]2eg_{\pi qq}^Wg_{\pi qq}\int {d^4p\over
(2\pi)^4}{(k_b+k_a)_\mu\over (k_b^2+m_\pi^2)(k_a^2+m_\pi^2)}\{{1\over
\gamma\cdot p-im}\gamma_5-\gamma_5{1\over \gamma\cdot p-im}\}$$
$$+2e{1\over 3}[-{2\over 3}]g_{\pi qq}^Wg_{\pi qq}\int {d^4k\over
(2\pi)^4}{1\over k^2+m_\pi^2}\{{1\over \gamma\cdot
p_b-im}+\gamma_\mu{1\over \gamma\cdot p_a-im}\gamma_5$$
$$-\gamma_5{1\over \gamma\cdot p_0-im}\gamma_\mu{1\over \gamma
\cdot p-im}\}.\eqno(2.6)$$
Here $k_b=p_{out}-p,\, k_a=p_{in}-p$  in the first integral (Fig. 1a) and
$p_b=p_{out}-k,\,p_a=p_{in}-k$ in the second integral (Fig. 1b). The 
overall coefficients are those for $u$-quarks, whereas the overall 
coefficients
for the $d$-quarks are given in the square brackets [...]. The
amplitudes (2.6) contain the appropriate flavor factors (Table 1).

The current operator (2.6) satisfies the current conservation constant
$q_\mu j_\mu=0$ only by addition of the contributions to the current from
the self-energy diagrams in Fig. 2. These divergent terms do not contribute
to the
$q$-dependence of the form factor. The calculation of the anapole moment
therefore should be carried out by dropping these terms and enforcing
current conservation on the loop amplitudes (2.6) after subtraction of
the divergent unphysical pole terms. The current transversality
requirement may be imposed directly by replacement of the operators
$\gamma_\mu$ and $(k_a+k_b)_\mu$ in (2.6) by
$$\gamma_\mu\rightarrow \gamma_\mu-{\gamma\cdot q\over q^2}q_\mu
\eqno(2.7a)$$
$$k_{a_\mu}+k_{b_\mu}\rightarrow k_{a_\mu}+k_{b_\mu}-{q\cdot
(k_a+k_b)\over q^2}q_\mu.\eqno(2.7b)$$
This procedure is equivalent to projecting out the longitudinal
component from the final result.

Pions are assumed to decouple from constituent quarks at the chiral
symmetry restoration scale $\Lambda_\chi\sim 4\pi f_\pi\sim 1.2$ GeV.
The loop integrals should accordingly be cut off at or about that
momentum scale. This should be done so as to maintain the
transversality condition $q_\mu j_\mu=0$. If in the quark coupling
(second) term in (2.5) the opinion propagator is replaced by
$${1\over m_\pi^2+k^2}\rightarrow v(k^2)={1\over
m_\pi^2+k^2}({\Lambda^2-m_\pi^2\over \Lambda^2+k^2})^2,\eqno(2.8)$$
the transversality condition is maintained, provided that the product of
the pion propagators in the pion coupling (first) term in (2.5) is
replaced by \cite{Gross}:
$${1\over k_b^2+m_\pi^2}{1\over k_a^2+m_\pi^2}\rightarrow
{v(k_b^2)-v(k_a^2)\over k_a^2-k_b^2}.\eqno(2.9)$$

Once the anapole moments $F_A^u$ and $F_A^d$ of the $u-$ and $d$-quarks
respectively have been calculated from the expression (2.6), the
corresponding proton- and neutron anapole form factors are obtained by
the standard quark model expressions as
$${F_A^p\over m_N^2}={4\over 3}{F_A^u\over m^2}-{1\over 3}{F_A^d\over
m^2},\eqno(2.10a)$$
$${F_A^n\over m_N^2}={4\over 3}{F_A^d\over m^2}-{1\over 3}{F_A^u\over
m^2}.\eqno(2.10b)$$
The expressions for the anapole form factors of the proton and the
neutron are then found to be
$$F_A^p(q^2)={m_N^2\over m^2}\{F_\pi(q^2)+{2\over 3}F_q(q^2)\},
\eqno(2.11a)$$
$$F_A^n(q^2)={m_N^2\over m^2}\{F_\pi(q^2)+{7\over 3}
F_q(q^2).\eqno(2.11b)$$
Here $F_\pi$ and $F_q$ are the contributions to the anapole moment of
the $u$-quark from the pion and quark coupling loops (Figs. 1 a,b)
in (2.6):
$$F_\pi(q^2)={m^2\over q^2}\{{g_{\pi qq}^Wg_{\pi qq}\over
4\pi^2}\int_{0}^{1}dxx\int_{0}^{1} dy\{(log{H_2(\Lambda_\chi^2)\over
H_2(m_\pi^2)}-x{\Lambda_\chi^2-m_\pi^2\over H_2(\Lambda_\chi^2)})$$
$$-(...)_{q^2=0}\},\eqno(2.12a)$$
$$F_q(q^2)=-{m^2\over q^2}{g_{\pi qq}^Wg_{\pi qq}\over 12\pi^2}
\int_{0}^{1}dx(1-x)\int_{0}^{1}dy$$
$$\{([m^2(1-x^2)-q^2(1-x)^2y(1-y)]K_1(q^2)+log{H_1(\Lambda_\chi^2)\over
H_1(m_\pi^2)}-x{\Lambda_\chi^2-m_\pi^2\over H_1(\Lambda_\chi^2)})$$
$$-(...)_{q^2=0}\}.\eqno(2.12b)$$
Here the notation $-(...)_{q^2=0}$ indicates that the value of the
preceeding bracket at $q^2=0$ should be subtracted from it. This
subtraction takes into account the self energy terms. 

The auxiliary functions in (2.12) are defined as \cite{Gloz,Han1}
$$H_1(M^2)=M^2x+m^2(1-x)^2+q^2(1-x)^2y(1-y),\eqno(2.13a)$$
$$H_2(M^2)=M^2x+m^2(1-x)+q^2x^2y(1-y).\eqno(2.13b)$$
Finally the function $K_1(q^2)$ is defined as
$$K_1(q^2)={1\over H_1(m_\pi^2)}-{1\over
H_1(\Lambda_\chi^2)}-x{\Lambda_\chi^2-m_\pi^2\over H_1^2(
\Lambda_\chi^2)}.\eqno(2.14)$$

The calculated pion loop contributions to the anapole form factors of
the proton and the neutron are shown in Fig. 3 as functions of
momentum transfer. In the calculation the "standard" value for $f_\pi$
(2.2a) $-4.5\cdot 10^{-7}$ \cite{DDH} was used. The constituent quark
mass was set to $m=340$ MeV and the value of the cut-off $\Lambda_\chi$
was taken to be 1.2 GeV.

The numerical values for pion loop contributions to the anapole moments
of the proton and the neutron were found to be $F_A^p(0)=3.22
\cdot 10^{-8}$ and $F_A^n(0)=1.45\cdot 10^{-8}$. These values
should be added to those obtained from the pion exchange 
and polarization currents
calculated below. Note similarity between the calculated proton and
neutron anapole moments, which indicates that main component of the
pion loop contribution to the anapole moment is the isoscalar term, in
agreement with other findings \cite{Henley1,Maekawa}. 
The expression for the pion loop contribution
to the isoscalar combination of the anapole form factors here is
formally similar to that, which appears in the derivation of 
the loop contribution at the baryon level with only
nucleon intermediate states \cite{Henley1}, the only
difference being the replacement of quark masses and
coupling constants by the corresponding nucleon masses and
coupling constants. The expressions for the isovector
amplitudes is different however, which is natural, as the
quark level calculation takes into account all baryonic
intermediate states, and in particular the $\Delta_{33}$
resonance.

\section{Pion exchange and polarization currents}

\vspace{0.5cm}

{\bf a. Parity violating pion exchange currents}

\vspace{0.5cm}

The pion exchange current operator, which is generated by the pion-quark
couplings (2.1), (2.4) and the e.m. currents of the pions and
constituent quarks (2.3a) are illustrated diagrammatically in Fig. 4.
In the absence of vertex form factors 
the expressions for the parity violating contact and pion current
exchange current operators are
$$j^\pi_\mu(C)
={-ieg_{\pi qq}^Wg_{\pi qq}\over 2m}[\vec \tau^1\cdot \vec
\tau^2-\tau^1_3\tau^2_3]\{{\gamma_5^1\gamma_\mu^1\over
k_2^2+m_\pi^2}+{\gamma_5^2\gamma_\mu^2\over k_1^2+m_\pi^2}\},\eqno(3.1a)$$
$$j^\pi_\mu(\pi)=-eg_{\pi qq}^Wg_{\pi qq}[\vec \tau^1\cdot \vec \tau^2-
\tau_3^1\tau_3^2]{(\gamma_5^2-\gamma_5^1)(k_1-k_2)_\mu\over (k_1^2+m_\pi^2)
(k_2^2+m_\pi^2)}.\eqno(3.1b)$$
Here the Dirac equation has been invoked for the quarks. The 4-momentum
fractions imparted to the two quarks are denoted $k_1,\, k_2$
respectively, so that $q=k_1+k_2$.

The current conservation condition on the PV pion exchange current
$j_\mu^{ex}(C)+j_\mu^{ex}(\pi)$ is \cite{Blunden}:
$$q_\mu j_\mu^{ex}=[V_\pi^{PV}(p'_1,p'_2,p_1+q,\, p_2)j_0^1(p_1+q,p_1)$$
$$-j_0^1(p'_1,p_1-q)V_\pi^{PV}(p'_1-q,p'_2,p_1,p_2)]$$
$$+(1\leftrightarrow 2).\eqno(3.2)$$
Here $j_0^1$ is the charge density operator of a single quark and
$V_\pi^{PV}$ is
the parity violating pion exchange potential for constituent quarks. The
initial and final 4-momenta of the quark pair are denoted $p_1,p_2$ and
$p'_1,p'_2$ respectively. The expression for the PV $\pi$-exchange
interaction is
$$V_\pi^{PV}=-ig_{\pi qq}g^W_{\pi qq}(\vec\tau^1\times \vec
\tau^2)_3{\gamma_5^1-\gamma_5^2\over k^2+m_\pi^2}.\eqno(3.3)$$
A practical consequence of the continuity equation (3.2) is the
requirement that the exchange current and
exchange induced polarization current contributions combine
to the proper non-singular form (1.1).

The pion exchange Yukawa functions in the exchange current operators
(3.1) as well as in the potential (3.3) should be cut-off at the chiral
restoration scale as were the pion loops. Current conservation is
maintained if in the contact current operator (3.1a) and the potential
the Yukawa function $1/(k^2+m_\pi^2)$ is modified as in (2.8), and the
product of pion Yukawa functions in the pionic current (3.1b) is
replaced as in (2.9) \cite{Gross}.

The current conservation condition (3.2) ensures that the exchange
current operator compensates for the non-conservation of the single
quark e.m. current in the nucleon states, in the presence of a parity
violating pion exchange interaction between the constituent quarks. If
the negative parity component of the proton wave function is neglected,
transversality has to be enforced on the net exchange current
contribution in the calculation of the anapole moment. This procedure
will be implemented here, although the polarization current that is
induced by the negative parity component is also considered 
explicitly below.

For the calculation of the matrix elements of the pion exchange current
operator it is conducive to rewrite it in the spin representation. To
lowest order in the inverse quark masses the operators (3.1) combine to
the expression
$$\vec j^\pi=-e{g_{\pi qq}^Wg_{\pi qq}\over 2m}(\vec \tau^1\cdot \vec
\tau^2-\tau^1_3\tau_3^2)
\{{\vec\sigma^2\over k_1^2+m_\pi^2}+{\vec\sigma^1\over
k_2^2+m_\pi^2}$$
$$-{(\vec \sigma^1\cdot \vec k_1-\vec \sigma^2\cdot
\vec k_2)(\vec k_1-\vec
k_2)\over (k_1^2+m_\pi^2)(k_2^2+m_\pi^2)}\}.\eqno(3.4)$$

In addition to the parameters that appear in the calculation of the pion
loop contributions to the anapole moment the exchange current
contribution also depends on the proton wave function and thus on the
confinement scale. The orbital part of proton
wave function will here be described by 
the oscillator function (Table 2)
$$\psi(\vec r,\vec \rho)=({\omega\over \sqrt{\pi}})^3e^{-(r^2+\rho^2)
\omega^2/2},\eqno(3.5)$$
which appears translationally invariant quark models as well as
in the covariant integrable quark model for the baryons
developed in ref. \cite{Coester1,Coester}. 
Here $\vec r$ and $\vec \rho$ are
Jacobi coordinates for the 3-quark system. With a constant flavor-spin
hyperfine interaction the baryon spectrum is well described with the
value $\omega = 311$ MeV \cite{Coester}, whereas in recent a meson exchange
model for the spectrum $\omega=1240$ MeV \cite{Helminen}. The latter
value is most consistent with the present meson exchange model for the
anapole moment.

The orbital matrix element of the pion exchange contact current (3.1a)
may be written as
$$j_\mu(C)=-{eg_{\pi qq}^Wg_{\pi qq}\over 2m}{m_\pi\over 4\pi}(\vec
\tau^1\cdot \vec
\tau^2-\tau^1_3\tau_3^2)[i\gamma_5^1\gamma_\mu^1+i\gamma_5^2 
\gamma_\mu^2]M_\rho(q)M_r(q),\eqno(3.6)$$
where $M_\rho(q)$ and $M_r(q)$ are the orbital matrix elements:
$$M_\rho(q)=4\pi({\omega\over
\sqrt{\pi}})^3\int_{0}^{\infty}d\rho\rho^2j_0({q\rho\over \sqrt{6}})e^{-
\rho^2\omega^2},\eqno(3.7)$$
$$M_r(q)=4\pi({\omega\over
\sqrt{\pi}})^3\int_{0}^{\infty}drr^2j_0({qr\over\sqrt{2}})
y_0(m_\pi r\sqrt{2})e^{-r^2\omega^2}.\eqno(3.8)$$
In the latter integral $y_0(x)$ is a cut-off Yukawa function defined as
$$y_0(m_\pi r\sqrt{2})={e^{-m_\pi r\sqrt{2}}\over m_\pi
r\sqrt{2}}-({\Lambda\over m_\pi}){e^{-\Lambda_\chi r\sqrt{2}}\over
\Lambda_\chi r\sqrt{2}}-{\Lambda_\chi^2-m_\pi^2\over 2\Lambda_\chi
m_\pi}e^{-\Lambda_\chi r\sqrt{2}}.\eqno(3.9)$$

For proton and neutron states with spin $z$-component $+1/2$, the matrix
element of the spin flavor operator calculated
with the wave functions listed in Table 3 is
$$<(\vec\tau^1\cdot \vec\tau^2-\tau^1_3\tau_3^2)(i\gamma^1_5
\gamma^2_3+i\gamma_5^2\gamma^2_3)>\simeq <(\vec \tau^1\cdot \vec
\tau^2-\tau_3^1\tau_3^2)
(\sigma_3^1+\sigma_3^2)>={4\over 3}.\eqno(3.10)$$
The contribution of pion exchange contact current to the anapole moment
of the proton and the neutron then takes the form
$$F_{A,C}^p(q^2)=F_{A,C}^n(q^2)=
{2\over 3}{m_N^2\over q^2}{g_{\pi qq}^Wg_{\pi qq}\over
4\pi}{m_\pi\over
m}[M_\rho(q)M_r(q)-M_\rho(0)M_r(0)].\eqno(3.11)$$
The subtraction of the matrix element at $q=0$ is required, unless a
consistent calculation of the polarization current is performed, in
which case the matrix element at $q=0$ is cancelled by a corresponding
term in the latter. This point will be treated in detail
below.

The calculation of the matrix element of the pionic exchange current
operator (3.1b) is somewhat more complicated. The result for this
contribution to the anapole moment may be cast into the form
$$F_{A,\pi}^p(q^2)=F_{A,\pi}^n(q^2)=
-{m_N^2\over q^2}{g_{\pi qq}^Wg_{\pi qq}\over
\pi}{1\over 18m}M_\rho(q)\int_{0}^{1}dx\{(m^*_\pi(x)$$
$$4\pi({\omega\over
\sqrt{\pi}})^3\int_{0}^{\infty}drr^2e^{-r^2\omega^2}\{j_0(\xi)[2y_0(
m_\pi^*r\sqrt{2})-y_1(m_\pi^*r\sqrt{2})]$$
$$-j_2(\xi)[y_0(m_\pi^*r\sqrt{2})+y_1(m_\pi^*r\sqrt{2})]\}$$
$$-(...)_{q^2=0}\}.\eqno(3.12)$$
Here the variable $\xi$ is defined as
$$\xi=qr({1\over 2}-x)\sqrt{2},\eqno(3.13)$$
and the function $y_1(m_\pi^*r\sqrt{2})$ is defined as:
$$y_1(m_\pi^* r\sqrt{2})=e^{-m_\pi^* r\sqrt{2}}-({\Lambda^*\over m_\pi^*}
)e^{-\Lambda^*r\sqrt{2}}$$
$$+{1\over 2}{\Lambda^{*2}-m_\pi^{*2}\over m_\pi^*\Lambda^*}
(1-\Lambda^* r\sqrt{2})
e^{-\Lambda^*r\sqrt{2}}.\eqno(3.14)$$
The mass parameters $m_\pi^*$ and $\Lambda^*$ are functions of the
integration variable $x$:
$$m_\pi^*(x)=\sqrt{m_\pi^2+q^2x(1-x)},\eqno(3.15a)$$
$$\Lambda^*(x)=\sqrt{\Lambda^2+q^2x(1-x)}.\eqno(3.15b)$$
The calculated pion exchange current contributions to the anapole form
factors of the proton and the neutron are shown in Fig. 5
The exchange current contributions are shown both as
calculated with the oscillator parameters $\omega = 311$ MeV and 1240 MeV
in order to obtain an estimate of the theoretical uncertainty of the
calculated values. 
The calculated value of the pion exchange current contribution to
the anapole moment of the proton (and the neutron) is -1.80 $\cdot 10^{-8}$
with
$\omega=$ 311 MeV and  $-1.15\cdot 10^{-8}$ with $\omega$ =1240 MeV.
As pointed out above the larger oscillator parameter
is consistent with the meson exchange model for the baryon wave functions
in ref. \cite{Helminen,Graz}. While the wave function dependence
of the anapole moment contribution itself is thus not very strong  
the exchange current contribution to the anapole form factor
is very strong.

For both parameter values a tendency for cancellation between the
exchange current and the pion loop contributions to the anapole moment
is visible. This then suggests that the net values of the anapole
moments of the nucleons will be small and only of the order $\sim 10^{-8}$.
In the
present quark model calculation the exchange current contribution
to the anapole moment is purely isoscalar, but the pion loop contribution
is about 2
times larger in the case of the proton than in the case of the
neutron. This result differs from that found in ref. \cite{Henley1},
where the PV pion loop fluctuations of the proton were considered.
The isoscalar nature in that loop calculation is a consequence
of dominance of the contributions of the terms, in which the e.m.
coupling is to the intermediate pions over the contributions
from the terms in which the e.m. coupling is to the intermediate
nucleon.

\vspace{0.5cm}

{\bf b. Parity violating pion exchange induced polarization current}

\vspace{0.5cm}

The parity violating pion exchange interaction (3.3) induces a negative
parity component into the proton wave function. In view of the small
magnitude of the PV pion exchange potential, this component may be
calculated by means of first order perturbation theory as
$$\psi^-=-\sum_{n}{<n|V_\pi^{PV}|0>\over E_n-E_0}\psi_n^-,\eqno(3.16)$$
where $\psi_n^-$ are the wave 
functions that describe the negative parity
excitations of the nucleons, and $E_n-E_0$ are the corresponding
resonance excitation energies. 

The electromagnetic current operator of the constituent quarks:
$$\vec j_q=e({1\over 6}+{\tau_3\over 2})\{{\vec p\,'+\vec p\over
2m}+{i\over 2m}\vec \sigma\times \vec q\},\eqno(3.17)$$
will have non-vanishing matrix elements between the positive $(\psi^+)$
and negative $(\psi^-)$ parity components of the nucleon wave function,
which contribute to the anapole form factor of the nucleon. These
contributions are referred to as polarization currents \cite{MusolfT}.
The polarization current then takes the form
$$\vec j^{pol}=-\sum_{n}(\psi^+,\,\sum_{q}\vec j_q{<n|V_\pi^{PV}|0>\over
E_n-E_0}\,\psi_n^-)$$
$$\,-\sum_{n}(\psi^-_n\,{<0|V_\pi^{PV}|n>\over E_n-E_0},\, \sum_{q}\vec
j_q\quad\psi^+).\eqno(3.18)$$
This expression is illustrated schematically in Fig. 7. 

The only ${1\over 2}^-$ nucleon states in the $P$-shell of the baryons
are the $N(1535)$ and $N(1650)$ resonances. The former is a member of an
$S=1/2$ negative parity doublet and the latter a member of an $S={3\over
2}$ negative parity quartet, in the usual assignment \cite{GloRis}. With
this assignment only the $N(1535)$ resonance contributes to the sum
(3.18). As the $P$-shell lies $\sim 600$ MeV above the nucleon, it is
justified to truncate the sum in (3.18) after the $P$-shell states, in
view of the much larger energy denominators of the resonances in the $F$
and higher shells.

The required matrix element of the PV pion exchange interaction may be
evaluated using the oscillator wave functions of the covariant quark
model in ref. \cite{Coester}, which are listed in
Table 2. The only non-vanishing matrix elements are
$$<N(1535)^+,\, {1\over 2}|V_\pi^{PV}|N(939)^+,{1\over 2}>
=\mp i{\sqrt{2}\over 3}{g_{\pi qq}g_{\pi qq}^W\over
4\pi}{m_\pi^2\over m}{\cal M},\eqno(3.19)$$
where ${\cal M}$ is the radial matrix element
$${\cal M}=\sqrt{2}  \omega({\omega\over
\sqrt{\pi}})^3 4\pi\int_{0}^{\infty}drr^3\bar y_1(m_\pi
r\sqrt{2})e^{-r^2\omega^2}.\eqno(3.20)$$
In (3.19) the $-$ sign applies for protons, and the $+$ sign for
neutrons. The Yukawa function $\bar y_1$ is defined as
$$\bar y_1(m_\pi r\sqrt{2})=(1+1/m_\pi r\sqrt{2})
{e^{-m_\pi r\sqrt{2}}\over m_\pi r\sqrt{2}}
-({\Lambda_\chi\over m_\pi})^2 (1+1/\Lambda_\chi r\sqrt{2})
{e^{-\Lambda_\chi r\sqrt{2}}\over \Lambda_\chi r\sqrt{2}}$$
$$-{1\over 2}(({\Lambda_\chi\over m_\pi}^2-1)e^{-\Lambda_\chi r
\sqrt{2}}.\eqno(3.21)$$

Evaluation of the matrix element (3.18) of the convection current part
(3.17) of the single quark current operator yields the following
contribution to the anapole form factors of the proton and the neutron:
$$F_{A,conv}^p=F_{A,conv}^n=-\eta {m_N^2\over q^2}{1\over 3}
{g_{\pi qq} g_{\pi qq}^W\over 4\pi}{\cal M}
{m_\pi^2\over  m^2}
{\omega\over \Delta}({\omega\over \sqrt{\pi}})^3
4\pi \int_0^\infty d\rho\rho^2
e^{-\rho^2\omega^2}
[j_0(\sqrt{{2\over 3}}q\rho)-1].\eqno(3.22)$$
Here $\Delta$ is the energy dominator $\Delta=1535-939=596$ MeV. Note
that this is an isoscalar term, as required by current conservation with
the isoscalar pion exchange current operator considered above. 
The factor $\eta$ is a correction factor that has to be chosen
so that the continuity equation that links the exchange current
and polarization current corrections is satisfied. If all 
negative parity states in the sum over $n$ in (3.18) are taken
into account, this factor would be unity. In order that no 
spurious pole at $q^2=0$ occur in the sum of pion exchange current
and polarization current matrix elements appear in the present
perturbative calculation of the latter, the value for the
correction factor $\eta$ has to be chosen as
$$\eta={2\over 3}{m\Delta\over \omega^2}.\eqno(3.23)$$
If the oscillator parameter $\omega$ is taken to be
$\omega$ = 311 MeV \cite {Coester}, 
the numerical value for $\eta$ is 1.4,
which indicates that truncating the sum over $n$ in
(3.18) is fairly good approximation. In the case of the
larger value 1240 MeV \cite{Helminen}
for $\omega$ the expression (3.23)
gives $\eta$ = 0.09, which indicates that higher lying excited
states contribute significantly.

Once the correction factor $\eta$ is included in the
expression (3.22), there is in principle no need to subtract
the values of the matrix elements at $q=0$ from the
exchange (3.11), (3.12) and polarization current contributions
(3.22) to the anapole moment. 
Above the value of the matrix elements at $q=0$ have been subtracted
explicitly, as it gives an explicit indication that the pole
term at $q^2$ = 0 has to cancel out.

The spin current component of the single quark current operator (3.17)
gives rise to the following contribution to the anapole moment of the
proton and the neutron:
$$F_{A,spin}^{p,n}=\eta(1,-{1\over 3})
{2\over 27}
{m_N^2\over  m^2}{m_\pi^2
{\cal M}\over
\Delta}{g_{\pi qq}g_{\pi qq}^W\over 4\pi}
\omega({\omega\over \sqrt{\pi}})^3 4\pi\int_0^\infty
d\rho\rho^4 e^{-\rho^2\omega^2}
[j_0(\sqrt{{2\over 3}}q\rho)+j_2(\sqrt{{2\over 3}}q\rho)].
\eqno(3.24)$$
Here the term 1 in the first bracket on the r.h.s. applies in the case
of the proton and the term $-1/3$ in the case of the neutron. 

The pion exchange induced polarization current contributions to the 
anapole form factors of the neutron and the proton have been plotted along
with the corresponding pion exchange current contributions in 
Figs. 5 and 6. The
polarization current contribution
to the anapole moment of the nucleon depends strongly on the
wave function model. With the wave function parameter value
$\omega$ = 311 MeV the polarization current contribution
to the anapole moment of the proton is $-3.38\cdot 10^{-8}$
and with the parameter value $\omega$ = 1240 MeV it
is $-0.19\cdot 10^{-8}$. The corresponding contributions
to the anapole moment of the neutron are $-0.38\cdot 10^{-8}$
and $-0.02\cdot 10^{-8}$ respectively.

The net combination of pion exchange and polarization
current contributions to the anapole form factors of the
proton and the neutron are shown in Fig. 8. The
net calculated pionic contribution to the anapole
moment of the proton is -- including the pion loop
contribution - is $1.88\cdot 10^{-8}$ and that
to the anapole moment of the neutron is $0.28\cdot 10^{-8}$
(for $\omega$ = 1240 MeV). 

\section{Vector meson loop contributions to the anapole moment}

\subsection{$\rho$-meson loop contributions}

The parity violating $\rho$-nucleon coupling is by convention written in
terms of 3 different flavor coupling terms \cite{Holstein}
$${\cal L}=i\bar\psi_N\{h^0_\rho\vec\tau\cdot \vec
\rho_\mu+h_\rho^1\rho_\mu^3+h_\rho^2{3\tau_3\rho_\mu^3-\vec\tau\cdot \vec
\rho_\mu\over 2\sqrt{6}}\}\gamma_\mu\gamma_5\psi_N.\eqno(4.1)$$
Here $\psi_N$ is the nucleon and $\vec \rho_\mu$ the isovector field of
the $\rho$-meson. The coupling constants $h_\rho$ have been determined
phenomenologically only within wide uncertainly ranges, the
"recommended" values being $h_\rho^0=-30g_w,\, h_\rho^1=-0.5 g_w$ and
$h_\rho^2=-25 g_w \,\, (g_w=3.8 \cdot 10^{-8})$ \cite{DDH}. 
Standard quark model algebra then implies that the PV coupling of
$\rho$-meson to constituent quarks be
$${\cal L}=i\bar\psi\{h_{\rho qq}^0\vec \tau\cdot \vec \rho_\mu+h_{\rho
qq}^1\rho_\mu^3+h_{\rho qq}^2{3\tau_3\rho_\mu^3-\vec \tau\cdot \vec
\rho_\mu\over 2\sqrt{6}}\}\gamma_\mu\gamma_5 \psi,\eqno(4.2)$$
where $\psi$ represents the quark fields. The PV $\rho$-quark coupling
constants $h_{\rho qq}$ are determined by the corresponding
$\rho$-nucleon coupling constants as
$$h_{\rho qq}^0={3\over 5}h_\rho^0,\, h_{\rho qq}^1=h_\rho^1,\, h_{\rho
qq}^2={3\over 5}h_\rho^2.\eqno(4.3)$$

The PV $\rho$-meson loop contributions to anapole moments of the
constituent $u$ and $d$ quarks are illustrated diagrammatically in Fig. 
9. The flavor factors for the different flavor coupling terms
$h_\rho$ are listed in Table 1 for the different diagrams. The
calculation of these loop contributions require
the $\rho$-quark coupling Lagrangian
$${\cal L}_{\rho qq}=ig_{\rho qq}\bar \psi\gamma_\mu\vec \tau\cdot
\vec\rho_\mu\psi,\eqno(4.4)$$
where $g_{\rho qq}=g_{\rho NN}\simeq 2.6$ \cite{GEB} and the e.m. current
operator for the $\rho$-meson
$$j_\mu=\pm
ie\{\rho^\dagger_\nu\partial_\mu\rho_\nu-\rho_\nu^\dagger\partial_\nu
\rho_\mu\}.\eqno(4.5)$$

With these couplings and the flavor factors listed in Table 1 the
expressions for the $\rho$-meson loop amplitudes, for $u$ quarks, where
e.m. coupling is to the internal quark (Fig. 9a), takes the form
$$j_\mu(u)=-{1\over 3}(2h_{\rho qq}^1+{h_{\rho qq}^2\over
\sqrt{6}})e\, g_{\rho qq}\int {d^4k\over
(2\pi)^4}
{\delta_{\alpha\beta}+{k_\alpha k_\beta\over m_p^2}\over k^2+m_p^2}$$
$$\{\gamma_\alpha\gamma_5 {1\over \gamma \cdot p_b-im}\gamma_\mu{1\over
\gamma\cdot p_a-im}\gamma_\beta\gamma_5$$
$$+\gamma_\alpha{1\over \gamma\cdot p_b-im}\gamma_\mu{1\over \gamma\cdot
p_a-im}\gamma_\beta \gamma_5\}.\eqno(4.6)$$
Here the notation is the same as in eqn. (2.6). The same expression
applies for $d$-quarks provided that the flavor
factor $(2h_{\rho qq}^1+h_{\rho
qq}^2/\sqrt{6})/3$ with the PV $\rho$-quark couplings is replaced by the
corresponding factor $-(h_{\rho qq}^0+h_{\rho qq}^1/3-h_{\rho qq}^2/
\sqrt{6}).$

The expression for the contribution of the corresponding $\rho$-meson
loop to fluctuations to the $u$-quark current, illustrated in Fig. 9b,
where the e.m. coupling is to the $\rho$-meson is
$$j_\mu(u)=2(h_{\rho qq}^0-{1\over \sqrt{6}}h_{\rho qq}^2)e g_{\rho
qq}\int {d^4p\over (2\pi)^4}{1\over k_b^2+m_p^2}{1\over k_a^2+m_p^2}$$
$$\{(k_a+k_b)_\mu(\delta_{\alpha\beta}+{k_{a\alpha}k_{a\beta}\over
m_\rho^2})(\delta_{\beta\delta}+{k_{b\beta}k_{b\delta}\over m_\rho^2})$$
$$\{\gamma_\delta\gamma_5{1\over \gamma\cdot
p-im}\gamma_\alpha+\gamma_\delta{1\over \gamma\cdot p-im}\gamma_\alpha
\gamma_5\}$$
$$-\gamma_\delta\gamma_5(\delta_{\beta\delta}+{k_{b\delta}k_{b\beta}\over
m_\rho^2}){k_{a\beta}\over \gamma\cdot
p-im}(\delta_{\mu\alpha}+{k_{a\mu}k_{a\alpha}\over m_\rho^2})\gamma_\alpha$$
$$-\gamma_\delta(\delta_{\beta\delta}+{k_{b\delta}k_{b\beta}\over
m_\rho^2}){k_{a\beta}\over \gamma\cdot
p-im}(\delta_{\mu\alpha}+{k_{a\mu}k_{a\alpha}\over
m_\rho^2})\gamma_a\gamma_5$$
$$-\gamma_\alpha\gamma_5(\delta_{\mu\alpha}+{k_{b\mu}k_{b\alpha}\over
m_\rho^2}){k_{b\beta}\over \gamma\cdot
p-im}(\delta_{\beta\delta}+{k_{a\delta}k_{a\beta}\over m_\rho^2})
\gamma_\delta$$
$$-\gamma_\alpha(\delta_{\mu\alpha}+{k_{b\mu}k_{b\alpha}\over
m_\rho^2}){k_{b\beta}\over \gamma\cdot
p-im}(\delta_\beta\delta+{k_{a\delta}k_{a\beta}\over
m_\rho^2})\gamma_\delta\gamma_5\}.\eqno(4.7)$$
The corresponding operator for $\rho$-meson fluctuations of the
$d$-quark in this case is obtained simply by change of the overall sign.
The sum of the current operators (4.6) and (4.7) satisfy the
tranversality condition when added to the combination of
$\rho$-meson self energy
diagrams (cf. Fig. 2). As those only add a $q$-independent
infinite contribution, which
has to be subracted, they are dropped here out and the
transversality condition on the $\rho$-meson loop currents (4.1) and
(4.7) is enforced as in the case of the pion loop diagrams above.

The current operators (4.6) and (4.7) simplify significantly by leaving
out the terms that are inversely proportional to $m_\rho^2$ in the
numerator of the 
$\rho$-meson propagators. As those terms have been found to be of very
small numerical significance for $q^2<m_p^2$ in other calculations of
related type \cite{Han1,Han2}, they
are dropped here, especially 
as their effect is small in comparison to that
of the wide uncertainty in the PV $\rho$-meson coupling constants
\cite{DDH}. 

The contributions to the anapole moments of the $u$ and $d$ quarks from
the $\rho$-meson loops with the e.m. coupling to the internal quark line
(Fig. 9 a) and to the $\rho$-meson (Fig. 9 b), respectively, are
then
$$F_{A,q}^{u,d}(q^2)
=f^{u,d}{m^2\over q^2}{g_{\rho qq}\over 4\pi^2}\int_{0}^{1}
dx(1-x)\int_{0}^{1}dy$$
$$\{[m^2(1-x)^2-q^2x-q^2(1-x)^2y(1-y)]\bar K_1(q^2,m_\rho^2)$$
$$+log{H_1(\Lambda_\chi^2)\over
H_1(m_\rho^2)}-x{\Lambda_\chi^2-m_\rho^2 \over
H_1^2(m_\rho^2)}\}-\{...\}_{q^2=0},\eqno(4.8a)$$
$$F_{A,\rho}^{u,d}(q^2)=g^{u,d}{m^2\over q^2}{g_{\rho qq}\over 4
\pi^2}\int_{0}^{1}dxx\int_{0}^{1}dy$$
$$\{[2m^2x(1-x)+xq^2+2x^2q^2y(1-y)]\bar K_2(q^2,m_\rho^2)$$
$$+6log {H_2(\Lambda_\chi^2)\over
H_2(m_\rho^2)}-6x{\Lambda_\chi^2-m_\rho^2\over
H_2^2(m_\rho^2)}\}-\{...\}_{q^2=0}.\eqno(4.8b)$$
Here $f^{u,d}$ and $g^{u,d}$ are the flavor factors
$$f^u={2h_{\rho qq}^1\over 3}+{h^2_{\rho qq}\over \sqrt{6}},\quad
f^d=h_{\rho qq}^0+{h_{\rho qq}^1\over 3}-{h_{\rho qq}^2\over
\sqrt{6}},\eqno(4.9a)$$
$$g^u=h_{\rho qq}^0-{h_{\rho qq}^2\over 2\sqrt{6}},\quad g^d=-h_{\rho
qq}^0+{h_{\rho qq}^2\over 2\sqrt{6}}.\eqno(4.9b)$$
The functions $K_1(q^2)$ and $K_2(q^2)$ are defined as
(cf. ref. \cite{Han1})
$$\bar K_1(q^2,m_\rho^2)={1\over H_1(m_\rho^2)}-{1\over
H_1(\Lambda_\chi^2)}-x{\Lambda_\chi^2-m_\rho^2\over
H_1^2(\Lambda_\chi^2)},\eqno(4.10a)$$
$$\bar K_2(q^2,m_\rho^2)={1\over H_2(m_\pi^2)}-{1\over
H_2(\Lambda_\chi^2)}-x{\Lambda_\chi^2-m_\rho^2\over H_2^2(
\Lambda_\chi^2)}.\eqno(4.10b)$$

The $\rho$--meson loop contributions to the anapole moments of the proton
and the neutron may then be calculated using eqs. (2.10). In Fig. 10
the calculated anapole form factors of the proton
and the neutron are shown. The magnitudes 
are the similar to those given by the corresponding pion loop 
contributions, but the signs are opposite. In
the numerical calculations the "recommended" values \cite{DDH} for the
PV $\rho$--nucleon coupling constants were employed.
The calculated values of the $\rho$--meson loop contributions to the 
anapole moments of the proton and the neutron were found to be
$-0.44 \cdot 10^{-8}$ and $1.52\cdot 10^{-8}$ respectively.  

The $\rho-$meson loops here are those, which are induced by the
PV $\rho-$meson-constituent quark couplings, which in turn are implied
by the $\rho-$nucleon couplings (4.1). In addition to these,
there may also appear a PV $\rho\rho\gamma$ coupling, akin to
the anapole moment term in the e.m. current operator. That
coupling has been considered in ref. \cite{MusolfT} and would
generate further PV loop contributions to the constituent quark
current. These are conserved by the form of the coupling, and
inversely proportional to $m_\rho^2$. In view of the uncertain
value of the PV $\rho\rho\gamma$ coupling strength and as the 
terms of order $m_\rho^{-2}$ 
were dropped from the loop calculation above, we have for 
consistency not included these loop contributions in the
present calculation.

\subsection{$\omega$-meson loop contributions}

The standard expression for the parity violating $\omega$-nucleon
coupling is
$${\cal
L}=i\bar\psi_N\{h_\omega^0\omega_\mu+h_\omega^1\tau^3\omega_\mu\}
\gamma_\mu\gamma_5\psi_N.\eqno(4.11)$$
The "recommended values" for the PV $\omega$-nucleon coupling constants
are $h_\omega^0=-5g_w$ and $h_\omega^1=-3g_w$, with a large uncertainty
margin \cite{DDH,Holstein}. 

The corresponding PV coupling of $\omega$ mesons to constituent
quarks is
$${\cal
L}=i\bar\psi\{h_{\omega qq}^0\omega_\mu+
h_{\omega qq}^1\tau^3\omega_\mu\}
\gamma_\mu\gamma_5\psi.\eqno(4.12)$$
The $SU(6)$ quark model for the nucleon wave functions leads to the
following relations between the PV $\omega$-nucleon and the
corresponding PV $\omega$-quark coupling constants:
$$h_{\omega qq}^0={h_\omega^0\over 3},\quad h_{\omega qq}^1={3\over
5}h_{\omega}^1.\eqno(4.13)$$
We shall rely on these relations here.

For the calculation of 
the contributions from the PV $\omega$-meson loop
fluctuations of the constituent $u$ and $d$ quarks to their anapole
moments, the $\omega$-quark coupling
$${\cal L}=ig_{\omega qq}\bar\psi \gamma_\mu\omega_\mu\psi\eqno(4.14)$$
is employed.
The $\omega$-quark coupling constant determined from the corresponding
$\omega$-nucleon vector coupling constant $g_{\omega NN}$ is $g_{\omega
qq}=g_{\omega NN}/3$ by the quark model. The recent Nijmegen model
\cite{Rijk} for the nucleon-nucleon interaction determines $g_{\omega
NN}=10.35$, from which it follows that $g_{\omega qq}=3.45$. This
value is used here.

Because of the neutrality of the $\omega$ meson the e.m. field
only couples to the internal quark in the PV $\omega$ meson
loop contributions.
The expressions for these contributions to the anapole
moments of the $u$ and $d$ quarks, then take the form
$$F_{A,\omega}^{u,d}(q^2)=-{m^2\over q^2}(h_{\omega qq}^0\mp h_{\omega
qq}^1)\int_{0}^{1}dx(1-x)\int_{0}^{1} dy$$
$$\{[m^2(1-x^2)-q^2x-q^2(1-x)^2y(1-y)]\bar K_1(q^2,m_\omega^2)$$
$$+log{H_1(\Lambda_\chi^2)\over
H_1(m_\omega^2)}-x{\Lambda_\chi^2-m_\rho^2\over
H_1^2(m_\rho)}\}-\{...\}_{q^2=0}.\eqno(4.15)$$

The expressions for the $\omega$-loop contributions to the nucleon
anapole form factors are then given by the equations (2.10). 
In Fig. 10 
the numerical values for the $\omega$-loop contributions to the
anapole form factors of the proton and the neutron
are also shown. 
The numerical values for the $\omega$-loop contributions to
the anapole moments of the nucleons are found to be 
--0.097$\cdot 10^{-8}$ (proton) and 0.34$\cdot 10^{-8}$ (neutron)
(Table 4).
The $\omega$ meson exchange current
contributions of to the anapole moments are expected to be
very small in view of the neutrality of the $\omega-$meson
and the small values of the PV $\omega$ coupling
constants as compares to the corresponding $\rho-$meson
couplings..

\section{Vector meson exchange and polarization currents}

\vspace{0.5cm}

{\bf a. Parity violating $\rho$ meson exchange currents}

\vspace{0.5cm}

The PV $\rho$-meson exchange current operators, which are implied by the
$\rho$-quark couplings (4.12), (4.4) are illustrated diagrammatically in
Fig. 11. The contact current is customarily derived as a pair current
operator \cite{MusolfT,Blunden}, but may also be derived as a contact
current as in the case of the pion exchange current $j_\mu(C)$
(3.1a) above. To see this one may rewrite the 
the parity conserving $\rho$-quark coupling (4.4) as
$${\cal L}_{\rho qq}=g_{\rho qq}\bar\psi
((p'+p)_\mu-{1\over 2m}
\sigma_{\mu\nu}\partial_\nu)\vec \rho_\mu\cdot\vec\tau \psi.\eqno(5.1)$$
Minimal substitution of the e.m. vector potential $A_\mu$ in this
coupling yields the contact coupling
$${\cal L}_{\gamma \rho qq}=e{g_{\rho qq}\over 2m}\bar\psi
\{\sigma_{\mu\nu}A_\nu(\vec \rho_\mu\times \vec \tau)_3
-A_\mu(\vec\rho_\mu\cdot\vec\tau+\rho_{\mu3}\}\psi\eqno(5.2)$$
in analogy with (2.4) in the case of the pion.

The expressions for the $\rho$-meson contact and $\rho$-current exchange
current operators are found to be
$$j_\mu^\rho(C)=-ie{g_{\rho qq}\over 2 m}{1\over k_2^2+m_\rho^2}
\{(h_{\rho qq}^0-{h_{\rho qq}^2\over 2\sqrt{6}})
(\vec \tau^1\times\vec\tau^2)_3
\gamma_\nu^2\gamma^2_5\sigma_{\nu\mu}^1$$
$$[(h_{\rho qq}^0-{h_{\rho qq}^2\over 2\sqrt{6}})\vec \tau^1
\cdot\vec \tau^2
+(h_{\rho qq}^0-{h_{\rho qq}^2\over \sqrt{6}})\tau^2_3
+h_{\rho qq}^1(1+\tau^1_3)+{3\over 2\sqrt{6}}\tau^1_3\tau^2_3)]\}
+(1 \leftrightarrow 2),\eqno(5.3a)$$
$$j_\mu^{\rho}(\rho)=ieg_{\rho qq}(h_{\rho qq}^0-{h_{\rho qq}^2\over
2\sqrt{6}}){(\vec \tau^1\times \vec \tau^2)_3\over
(k_1^2+m_\rho^2)(k_2^2+m_\rho^2)}$$
$$\{(k_1-k_2)_\mu[\gamma_\nu^2\gamma_5^2\gamma^1_\nu
+\gamma_\nu^1\gamma_5^1\gamma_\nu^2]$$
$$-\gamma_\nu^2\gamma_5^2 k_{1\nu}\gamma_\mu^1
+\gamma_\nu^1\gamma_5^1
k_{2\nu}\gamma_\mu^2+\gamma_\mu^2\gamma_5^2\gamma_\nu^1 
k_{2\nu}-\gamma_\mu^1\gamma_5^1\gamma_\nu^2 k_{1\nu}\}
.\eqno(5.3b)$$

In the spin representation the sum of the local parts of the
$\rho$-meson exchange current operators
(5.3), to lowest order in $1/m^2$, reduce to the expression
$$\vec j^\rho=e(h_{\rho qq}^0 -{h_{\rho qq}^2\over 2\sqrt{6}}){g_{\rho
qq}\over 2m}(\vec \tau^1\times \vec \tau^2)_3$$
$$\{(\vec \sigma^1\times\vec \sigma^2)\{{1\over k_1^2+m_\rho^2}+{1\over
k_2^2+m_\rho^2}\}-{1\over (k_1^2+m_\rho^2)(k_2^2+m_\rho^2)}$$
$$\{(\vec \sigma^1\times\vec \sigma^2)\cdot (\vec k_1-\vec k_2)
(\vec k_1-\vec k_2)+\vec\sigma^2\cdot \vec k_1
\vec\sigma^1\times \vec k_1-\sigma^1\cdot k_2
\vec\sigma^2\times\vec k_2
-\vec\sigma^2\vec\sigma^1\cdot \vec k_1\times\vec k_2
+\vec\sigma^1 \vec\sigma^2\cdot\vec k_1\times\vec k_2
\}\}$$
$$+e{g_{\rho qq}\over 2 m}\{{\vec\sigma^2\over k_2^2+m_\rho^2}
[(h_{\rho qq}^0-{h_{\rho qq}^2\over 2\sqrt{6}})\vec \tau^1
\cdot\vec \tau^2
+(h_{\rho qq}^0-{h_{\rho qq}^2\over \sqrt{6}})\tau^2_3
+h_{\rho qq}^1(1+\tau^1_3)+{3\over 2\sqrt{6}}\tau^1_3\tau^2_3)]$$
$$+{\vec\sigma^1\over k_1^2+m_\rho^2}
[(h_{\rho qq}^0-{h_{\rho qq}^2\over 2\sqrt{6}})\vec \tau^1
\cdot\vec \tau^2
+(h_{\rho qq}^0-{h_{\rho qq}^2\over \sqrt{6}})\tau^1_3
+h_{\rho qq}^1(1+\tau^2_3)+{3\over 2\sqrt{6}}\tau^1_3\tau^2_3)]
.\eqno(5.4)$$

Apart from the coupling constants, this PV $\rho$-meson exchange current
operator has a formal similarity to that of the corresponding PV pion
exchange current operator (3.4). In contrast to the latter the
$\rho$-meson exchange current operator is an isovector operator.

Comparison of the expressions (3.4) and (5.4) for the PV $\pi$ and
$\rho$ meson exchange current operator makes it possible to derive the
$\rho$-meson exchange contribution to the anapole moments of the proton
and the neutron by analogy. The required spin-isospin
matrix elements are given in Table 3.

The contribution of the seagull current (5.3a) to the anapole
moment of the proton and the neutron becomes  
becomes
$$F_{A,C}(q)=4{m_N^2\over q^2}h_C(p,n){g_{\rho qq}\over 4\pi}$$
$${m_\rho\over m}[M_\rho(q)\bar M_r(q)-M_\rho(0)\bar M_r(0)].
\eqno(5.5)$$
Here the coefficients $h_C(p,n)$ are defined as
$$h_C(p,n)=\pm(h_{\rho qq}^0-{h_{\rho qq}^2\over 2\sqrt{6}})
-{1\over 2}(h_{\rho qq}^0-{h_{\rho qq}^2\over 2\sqrt{6}})
\mp{5\over 12}(h_{\rho qq}^0-{h_{\rho qq}^2\over \sqrt{6}})
\mp {h^1_{\rho qq}\over 4}+{1\over 12}h^1_{\rho qq}-{h^2_{\rho qq}
\over 8\sqrt{6}}. \eqno(5.6)$$
The upper and lower signs in this expression applies to
protons and neutrons respectively.
The function $M_\rho(0)$ is defined in (3.7), and the matrix element
$\bar M_r(q)$ is defined as
$$\bar M_r(q)=4\pi({\omega\over
\sqrt{\pi}})^3\int_{0}^{\infty}drr^2j_0({qr\over
\sqrt{2}})y_0(m_\rho\sqrt{2}r)e^{-r^2\omega^2}\eqno(5.7)$$
The cut-off Yukawa function $y_0$
is defined in (3.9), the $\rho$-meson mass
here having been substituted for the pion mass.

The contribution to the anapole moment from the PV $\rho$-meson exchange
current operator (5.3b) (second term in (5.4)), where the e.m. coupling
is to the $\rho$-meson field, takes the form (cf. (3.12))
$$F_{A,\rho}^p(q)=-F_{A,n}^n(q)=-{m_N^2\over q^2}(h_{\rho qq}^0-{h_{\rho
qq}^2\over 2\sqrt{6}}){g_{\rho qq}\over 4\pi}{1\over m}M_\rho(q)$$
$$\int_{0}^{1}dx\{m_\rho^*(x)4\pi({\omega\over
\sqrt{\pi}})^3\int_{0}^{\infty}drr^2e^{-r^2\omega^2}$$
$$\{2j_0(\xi)[2y_0(m_\rho^*r\sqrt{2})-(1-{\vec q\,^2
\over 8m_\rho^{*2}})y_1(m_\rho^*r\sqrt{2})]$$
$$-j_2(\xi)[y_0(m_\rho^* r\sqrt{2})+y_1(m_\rho^*
r\sqrt{2})]-(...)_{q=0}\}.\eqno(5.8)$$
Here the notation is the same as used in the expression (3.11). The mass
function $m_\rho^*(x)$ is defined as (cf. 3.15a)
$$m_\rho^*(x)=\sqrt{m_\rho^2+q^2x(1-x)}.\eqno(5.9)$$

The net calculated $\rho$-meson exchange current contributions
$F_A^\rho(q,C)+F_A^\rho(q,\rho)$ to the anapole form factors of the
proton and the neutron are 
shown in Figs. 12 and 13 respectively for two different values of the 
wave function parameter $\omega$. With $\omega$ = 311 MeV these
contributions to
the anapole moments of the proton and the neutron are
$-1.18\cdot 10^{-8}$ and $-0.85\cdot 10^{-8}$ respectively.
With the larger value $\omega = 1240$ MeV, which is consistent
with the Hamiltonian model for the baryon spectrum in ref. 
\cite{Helminen,Graz}, the corresponding values are
$-1.79\cdot 10^{-8}$ and $0.73\cdot 10^{-8}$ respectively.

In ref. \cite{MusolfT} another type of PV $\rho-$meson
exchange current was considered. That exchange current arises
from the PV $\rho\rho\gamma$ coupling, which corresponds to
the anapole coupling of the $\gamma$ to the nucleons. This
isoscalar exchange current is conserved by construction,
and inversely proportional to $m_\rho^2$. Because of the
uncertain value of the PV $\rho\rho\gamma$ coupling
strength and as the terms of
order $m_\rho^{-2}$ have not been included in the calculation above,
this isoscalar $\rho-$ meson exchange current has not
been included here for reasons of consistency.

\vspace{0.5cm}

{\bf b. Parity violating $\rho$ exchange induced polarization current}

\vspace{0.5cm}

The parity violating $\rho$ meson exchange interaction between
constituent quarks contributes to the negative parity component of the
proton wave function. This interaction takes the form \cite{Holstein}
$$V_\rho^{PV}=-{g_{\rho qq}\over 4\pi}\{h_{\rho qq}^0\vec\tau^1\cdot
\vec \tau^2+{1\over 2}h_{\rho qq}^1(\tau^1_3+\tau^2_3)$$
$$+{1\over 2\sqrt{6}}h_{\rho qq}^2(3\tau_3^1\tau_3^2-\vec \tau^1\cdot
\vec \tau^2)\}$$
$$\{(\vec \sigma^1-\vec \sigma^2)\cdot\{{\vec p_1-\vec p_2\over 2m},\,
f(r)\}_+\}
+i\vec\sigma^1\times\vec\sigma^2\cdot[{\vec p_1-\vec p_2\over 2 m},
f(r)]\}$$
$$+{g_{\rho qq}\over 4\pi}{h^1_{\rho qq}\over 2}(\tau^1_3
-\tau^2_3)(\vec\sigma^1+\vec\sigma^2)\cdot[{\vec p_1
-\vec p_2\over 2 m},f(r)].\eqno(5.10)$$
The radial Yukawa function $f(r)$ is defined as
$$f(r)={e^{-m_\rho r}\over r}-{e^{-\Lambda_\chi r}\over r}-
{m_\rho^2\over
2\Lambda_\chi}({\Lambda_\chi^2\over m_\rho^2}-1)e^{-\Lambda_\chi
r}.\eqno(5.11)$$
Here $r=|\vec r_1-\vec r_2|$.
The $\rho$-exchange induced PV polarization current is then given by the
general expression (3.18) with $V_\rho^{PV}$ substituted in place of
$V_\pi^{PV}$. In order to be consistent with the approximate
treatment of the $\rho-$meson exchange current operators above,
where only the local part of the operator was retained, 
the $\rho-$meson exchange induced polarization
contribution is calculated here using only the 
local (commutator) terms in the
$\rho-$meson exchange interaction.

If the sum over negative parity states is approximated by the
contributions from the lowest lying negative parity resonances, $N(1535)$
and $\Delta(1620)$,
the only matrix elements that are required are
$$<N(1535)^+,\,{1\over 2}|V_\rho^{PV}|N(939)^+,\, {1\over 2}>
= i\sqrt{2}{g_{\rho qq}\over
4\pi}{m_\rho^2\over m}({h_{\rho qq}^0\over 2}
\mp {h_{\rho qq}^1\over 3})
{\cal M}_\rho,\eqno(5.12a)$$
$$<\Delta(1620)^+,{1\over 2}|V_\rho^{PV}|N(939)^+,{1\over 2}>=
- 2i\sqrt{2}{g_{\rho qq}\over 4\pi}{m_\rho^2\over m}
(\pm{h_{\rho qq}^1\over 6}+{h_{\rho qq}^2\over
2\sqrt{6}}){\cal M}_\rho^a
.\eqno(5.12b)$$
Here the orbital matrix element ${\cal M}_\rho$
is defined as
$${\cal M}_\rho=\sqrt{2}\omega({\omega\over
\sqrt{\pi}})^34\pi\int_{0}^{\infty}drr^3e^{-r^2\omega^2}\bar 
y_1(m_\rho
r\sqrt{2}),\eqno(5.13)$$
Here the wave function models in Table 2 has again been
employed. The
function $\bar y_1$ is defined as in (3.12). 

Evaluation of the matrix element (3.18) of the convection current part
of the single quark current (3.17), with $V_\pi^{PV}$ replaced by
$V_\rho^{PV}$, with the sum over $n$ truncated to include only the
contributions from the $N(1535)$ and $\Delta(1620)$ negative parity
resonances then leads to the following anapole form factor
contributions to the proton and the neutron:
$$F_{A,conv}^p=-F_{A,conv}^n=\eta{m_N^2\over q^2}
{g_{\rho qq}\over 4\pi}
{m_\rho^2\over m^2}
\{{1\over \Delta_1}({h^0_{\rho qq}\over 2}
\mp {1\over 3}h_{\rho qq}^1)
{\cal M}_\rho$$
$$+{2\over \Delta_2}((\mp{h_{\rho qq}^1\over 6}
+{h_{\rho qq}^2\over 2\sqrt{6}})
{\cal M}_\rho
\}\omega({\omega\over
\sqrt{\pi}})^34\pi\int_{0}^{\infty}d\rho\rho^2e^{-\rho^2\omega^2}[j_0
(\sqrt{{2\over 3}}q\rho)-1].\eqno(5.14)$$
Here $\Delta_1$ and $\Delta_2$ are the mass differences
$m[N(1535)]-m_N=596$ MeV and $m[\Delta(1620)-m_N]=681$ MeV respectively.
The same renormalization correction factor $\eta$ (3.23) as
was  employed in the expression for the pion exchange induced
parity violating polarization current has been used here. The required
renormalization constant was derived explicitly from the continuity
equation in the case of pion exchange. It cannot be calculated in the
same way in the case of $\rho$-meson exchange, because the PV
$\rho$-coupling (4.1) does not satisfy the required transversality
condition. \\

The spin current component of the single quark current operator (3.17)
gives the following $\rho$-meson exchange induced polarization current
contribution to the anapole form factors of the proton and the neutron:
$$F_{A,spin}^{p,n}=-\eta(1,-{1\over 3})
{2\over 9}{g_{\rho qq}\over 4\pi}{m_N^2 m_\rho^2\over m^2}
\{{1\over \Delta_1}({h^0_{\rho qq}  \over 2}
\mp{1\over 3}h_{\rho qq}^1)
{\cal M}_\rho$$
$$+{2\over 3\Delta_2}((\pm{h_{\rho qq}^1\over 6}
+{h_{\rho qq}^2\over 2\sqrt{6}})
{\cal M}_\rho\}$$
$$\omega({\omega\over
\sqrt{\pi}})^34\pi\int_{0}^{\infty}d\rho\rho^4e^{-\rho^2\omega^2}
[j_0(\sqrt{{2\over 3}}q\rho)+j_2(\sqrt{{2\over
3}}q\rho)].\eqno(5.15)$$
Here the term 1 in the bracket $(1, -1/3)$ on the r.h.s. applies in the
case of the proton and the term $-1/3$ in the case of the neutron.

The $\rho-$meson exchange induced contributions to the anapole
form factor of the proton is shown in Fig. 
12 as obtained with
two different values for the wave function parameter $\omega$.
The corresponding contributions to the anapole form factor
of the neutron are shown in Fig. 13. For $\omega =$ 1240 MeV the
$\rho$-exchange induced polarization current contribution to the anapole
moment of the proton is $+0.088 \cdot 10^{-8}$ and that of the neutron
is $-0.047\cdot 10^{-8}$ (Table 4).

\section{Meson loops with transition couplings}

In addition to the pion and vector meson loop contributions to the
anapole form factors, loop contributions with e.m. pseudoscalar-vector
meson transition couplings also contribute to the anapole form factors.
These are illustrated by the Feynman diagrams in Fig. 14.

The $\rho\rightarrow \pi\gamma$ transition coupling may be descibed by
the current matrix element
$$<\pi^a(k')|j_\mu(0)|\rho^b_\nu(k)>=-i{g_{\rho \pi\gamma}\over
m_\rho}\epsilon_{\mu\nu\lambda\sigma} k_\lambda k'_\sigma
\delta^{ab}.\eqno(6.1)$$
The $\rho\pi\gamma$ coupling constant may be calculated from the known
radiative widths of the $\rho$-mesons:
$\Gamma(\rho^\pm\rightarrow \pi^\pm\gamma)=68$ keV,
$\Gamma(\rho^0\rightarrow \pi^0\gamma)=119$ keV. These yield
$g_{\rho^\pm\pi^\pm\gamma}=0.57$ and $g_{\rho^0\pi^0\gamma}=0.75$.

The corresponding transition current matrix element for the
$\omega\rightarrow \pi^0\gamma$ transition is
$$<\pi^0(k')|j_\mu(0)|\omega_\nu(k)>=-i{g_{\omega \pi\gamma}\over
m_\omega}\epsilon_{\mu\nu\lambda\sigma}k_\lambda k'_\sigma.\eqno(6.2)$$
From the radiative width of the $\omega$-meson:
$\Gamma(\omega\rightarrow \pi^0\gamma)=717$ MeV one obtains
$g_{\omega\pi\gamma}=1.84$. 

The current operators that describe the $\rho\rightarrow \pi\gamma$
transition loops on $u$- and $d$-quarks in Fig. 9, where one of the
hadronic vertices is parity violating takes the form
$$j_\mu^{u,d}=-i{G^{u,d}\over m_\rho}\int{d^up\over
(2\pi)^4}\{{\epsilon_{\mu\nu\lambda\sigma}k_{a\lambda}k_{b\sigma}\over
(k_b^2+m_\pi^2)(k_a^2+m_\rho^2)}{1\over \gamma\cdot p-im}\gamma_\nu$$
$$+\gamma_\nu{1\over \gamma
p-im}{\epsilon_{\mu\nu\lambda\sigma}k_{a\lambda}k_{b\sigma}\over
(k_b^2+m_\rho^2)(k_a^2+m_\pi^2)}\}.\eqno(6.3)$$
Here the coupling constant combinations $G^u$ and $G^d$ have been
defined as
$$G^{u,d}=2g_{\pi qq}^Wg_{\rho qq}g_{\rho^\pm\pi^\pm\gamma}+g_{\pi
qq}[2h^0_{\rho qq}g_{\rho^\pm\pi^\pm\gamma}$$
$$+(h_{\rho qq}^0\pm h_{\rho qq}^1+{h_{\rho qq}^2\over
\sqrt{6}})g_{\rho^0\pi^0\gamma}].\eqno(6.4)$$
The only difference between $G^u$ and $G^d$ is in the sign of the
$h_{\rho qq}^1$ term, which is positive in the case of $G^4$ and
negative in the case of $G^d$.

Because of the presence of the Levi-Civita symbol in the
current expressions (6.3), the apparent ultraviolet
divergences drop out. Similarly the terms $k_\alpha k_\beta/m_\rho^2$ in
the vector meson propagators drop out.
In line 
with the cutting off of the loop integrals of the pion and vector 
meson loop
contributions above at the chiral symmetry restoration scale,
a similar cutoff is applied here by
insertion of a form factor,
$$F(k_\pi^2,k_\rho^2)={\Lambda_\chi^2-m_\pi^2\over
\Lambda_\chi^2+k_\pi^2}{\Lambda_\chi^2-m_\rho^2\over
\Lambda_\chi^2+k_\rho^2},\eqno(6.5)$$
in the integrands, where $k_\pi$ and $k_\rho$ denote the 4-momenta of
the $\pi$ and $\rho$ mesons.

To reduce expressions (6.3) to standard form (1.1), it is convenient to
employ the relation
$$\epsilon_{\mu\nu\lambda\sigma}(p'+p)_\lambda q_\sigma \gamma_\nu
=q^2\gamma_\mu\gamma_5-2im \gamma_5 q_\mu,\eqno(6.6)$$
where $q_\mu=p'_\mu-p_\mu$. the PV $\rho-\pi$ loop contributions to the
anapole moments of the $u$ and $d$ quarks then become
$$F_{A,\rho\pi}^{u,d}(q^2)={G^{u,d}\over 16\pi^2}{m\over
m_\rho}\int_{0}^{1}dxx\int_{0}^{1}dy(1-x)[1+x(1-2y)]$$
$$\{{1\over G(m_\rho,m_\pi)}-{1\over G(m_\rho,\Lambda_\chi)}-{1\over
G(\Lambda_\chi,m_\pi)}+{1\over G(\Lambda_\chi,\Lambda_\chi)}\}.\eqno(6.6)$$
Here the function $G(m_1,m_2)$ has been defined as
$$G(m_1,m_2)=m^2(1-x)^2+m_1^2x(1-y)+m_2'^2xy+q^2x^2y(1-y).\eqno(6.7)$$
The contribution from the PV loops that involve the $\omega\rightarrow
\pi\gamma$ transition in the meson line may be calculated by the same
method. This contribution obtains no contribution from the PV
$\pi$-quark coupling (2.1a), which only contributes to loop amplitudes
with charged pions. The PV $\pi\gamma$ transition loops leads to the
following anapole form factors of the $u$ and $d$ quarks:
$$F_{A,\omega\pi}^{u,d}(q^2)={H^{u,d}\over 16\pi^2}{m\over
m_\omega}\int_{0}^{1}dxx\int_{0}^{1}dy(1-x)[1+x(1-2y)]$$
$$\{{1\over G(m_\omega,m_\pi)}-{1\over G(m_\omega,\Lambda_\chi)}-{1\over
G(\Lambda_\chi,m_\omega)}+{1\over
G(\Lambda_\chi,\Lambda_\chi)}\}.\eqno(6.8)$$
Here a cut-off factor of the form (6.5) with $m_\rho$ replaced by
$m_\omega$ has been included in the expression. The factors $H^{u,d}$
represent the coupling constant combinations
$$H^{u,d}=-g_{\pi qq}(h_{\omega qq}^0\pm h_{\omega qq}^1)g_{\omega
\pi\gamma}.\eqno(6.9)$$ 

The $\rho\pi$ and $\omega\pi$ loop contributions to the anapole moment
of the proton, as calculated using the expressions (2.10) have 
been plotted in Fig. 15.
The contribution from the $\rho\pi$ loops is the larger
one because of the larger PV $\rho-$quark coupling strengths.
These contribute $-0.67\cdot 10^{-8}$ and $-0.65\cdot 10^{-8}$
to the anapole moments of the proton and the neutron (Table 4).
The corresponding contributions from the $\omega\pi$ loops to
the anapole moments of the proton and the neutron are
$0.12\cdot 10^{-8}$ and $-0.035\cdot 10^{-8}$ respectively.

The $\rho\rightarrow \pi\gamma$ and $\omega \rightarrow \pi\gamma$
transition couplings also give rise to parity violating exchange current
operators. These are illustrated
diagrammatically in Fig. 16. The corresponding 2-quark currents may be
derived by using the meson-nucleon couplings in sections 2 and 4 and the
$\rho\rightarrow \pi\gamma$ and $\omega\rightarrow \pi\gamma$ transition
couplings (6.1) and (6.2).

These exchange current operators do not however contribute to the
anapole form factors of the nucleon. The exchange current operator that
involves the PV pion-quark coupling (2.1a) is spin-independent, and
vanishes in nucleon states because of the antisymmetric Levi-Civita
symbol in the e.m. pion-vector meson transition couplings (6.1) and
(6.2). The exchange currents that involve the PV vector meson couplings
(4.2) (4.11) have no local component, which would contribute to the
anapole form factor.

\section{Discussion}

The calculated net anapole form factors of the proton and
the neutron are shown in Fig. 17. These results take into
account all the loop
and exchange current contributions calculated above.
As a satisfactory description of the nucleon radii with
the constituent quark model of ref. \cite{Graz} requires
that a form factor be associated with the constituent
quarks \cite{Helminen}, the calculated anapole form factors have been
multiplied by the corresponding form factor in Fig. 17.
The form of the constituent quark form factor used
was $exp(-r^2q^2/6)$, where the mean square radius
value is $r^2=0.133$ fm$^2$ \cite{Helminen}. The calculated
anapole moments are also shown without this quark form factor
modification.

The corresponding contributions to the anapole moments of the
proton and the neutron are listed in Table 4.
The calculated value of the anapole moment of the proton
is $ -0.90\cdot10^{-8}$, the magnitude of whichwhich
is considerably smaller than what is obtained by considering
only the pionic contributions.
The main difference from the 
PV pion-nucleon loop estimate in
ref. \cite{Henley1}
and the chiral perturbation theory estimate in ref. \cite{Maekawa}
is the appearance of exchange and polarization current
contributions at the quark level.
The present calculation suggests that the net mesonic
anapole moment is far too small to lead to any notable 
correction to the
extraction of the strangeness magnetic moment in the SAMPLE
experiment \cite{SAMPLE1,SAMPLE2}. The net mesonic contribution to the
anapole moment of the neutron is $0.68 \cdot 10^{-8}$.

The calculated anapole moment has a wide $\sim 100\%$ theoretical
uncertainty, which mainly arises from the large uncertainty
in the PV meson-nucleon
couplings \cite{DDH}, which are used as input parameters in the
calculation. The loop amplitudes have a cut-off sensitivity, but this is
fairly small, once the cut-off is taken to be of the order of the chiral
symmetry restoration scale. The exchange current and polarization
current contributions have an additional
dependence on the baryon wave function model, which to some
extent 
is limited by requirement of consistency with the calculated baryon
spectrum \cite{Graz}.

A main qualitative result of the present calculation is that 
at the quark level there is a tendency for cancellation between
the loop and exchange current contributions
and the tendency for cancellation between pionic
and vector meson contributions.

\vspace{2cm}

{\bf Acknowledgement}
\vspace{0.5cm}

I thank Professor R. D. McKeown for his hospitality at the W. K.
Kellogg Radiation Laboratory of the California Institute of Technology
during the completion of this work. Research supported by the Academy of
Finland under contract 43982.

\newpage

\centerline{\bf Table 1}

\vspace{0.5cm}

Flavor factors in the loop integrals, when the parity conserving
coupling preceeds the parity violating one. The factors with an asterisk
change sign when the ordering of the couplings is reversed.

\vspace{0.5cm}

\begin{center}
\begin{tabular}{|l|c|} \hline
 & \\
$u\rightarrow \pi^+d\rightarrow u$ & $\quad 2i^*\quad$\\

$u\rightarrow \pi^0 u\rightarrow u$ & $\quad 0\quad$\\

$d\rightarrow \pi^-u\rightarrow d$ & $\quad 2i^*\quad$\\

$d\rightarrow \pi^0 d\rightarrow d$ & $\quad 0\quad$\\ 
 & \\ \hline
 & \\
$u\rightarrow \rho^+d\rightarrow u$ & $2 \quad (h_{\rho qq}^0)$\\

& $0 \quad(h_{\rho qq}^1)$\\

& $-1/\sqrt{6}\quad(h_{\rho qq}^2)$\\
 & \\  
$u\rightarrow \rho^0u\rightarrow u$ & 1 $\quad(h_{\rho qq}^0)$\\

 & $1\quad(h_{\rho qq}^1)$\\

& $1/\sqrt{6}\quad (h_{\rho qq}^2)$\\
 & \\
$d\rightarrow \rho^-u\rightarrow d$ & $2\quad (h_{\rho qq}^0)$\\

 & $-1/\sqrt{6}\quad (h_{\rho qq}^2)$\\
 & \\
$d\rightarrow \rho^0d\rightarrow d$ & $1\quad (h_{\rho qq}^0)$\\

& $-1\quad (h_{\rho qq}^1)$\\

& $1/\sqrt{6}\quad (h_{\rho qq}^2)$\\
 & \\
$u\rightarrow \rho^0u\rightarrow u$ & $\quad 1(h_{\omega qq}^0)$\\

& $\quad 1(h_{\omega qq}^1)$\\
 & \\
$d\rightarrow \rho^0u\rightarrow u$ & $\quad 1(h_{\omega qq}^0)$\\

 & $\quad -1(h_{\omega qq}^1)$\\  
 & \\ \hline
\end{tabular}
\end{center}

\newpage

\centerline{\bf Table 2}

\vspace{0.5cm}

Explicit wave functions for the ${1\over 2}^+$ and ${1\over 2}^-$ 
non-strange baryon states in the lowest $S$- and $P$-shells. The 
functions $\varphi_{nlm}$ are harmonic oscillator wave functions. The 
subscripts $\pm$ on the spin-isospin states denote the Yamanouchi symbols 
(112) and (121) respectively \cite{Coester1}.

\vspace{0.5cm}

\begin{center}
\begin{tabular}{|c|l|} \hline
 & \\
$p,n$ & ${1\over \sqrt{2}}\varphi_{000}(\vec \rho)\varphi_{000}(\vec r)
\{|{1\over 2},t>_+|{1\over 2},s>_++|{1\over 2},t>_-|{1\over 2}
s>_-\}$\\
 & \\ \hline
 & \\
$N(1535)$ & ${1\over 2}\sum_{m\sigma}(1,{1\over 2},m,\sigma|{1\over 2}
s)\{\varphi_{01m}(\vec \rho)\varphi_{000}(\vec r)$\\
 & \\
 & $[{1\over 2},t>_+|{1\over 2}\sigma>_+-|{1\over 2},t>_-|{1\over 2},
\sigma>_-]$\\
 &  \\
 & $+\varphi_{000}(\vec \rho)\varphi_{01m}(\vec r)[|{1\over 2},t>_+|
{1\over 2},\sigma>_-+|{1\over 2},t>_-|{1\over 2},\sigma>_+]\}$ \\
 & \\ \hline
  & \\
$\Delta(1620)$ & ${1\over \sqrt{2}}\sum_{m\sigma}(1,{1\over 2},m,\sigma|
{1\over 2},s)\{\varphi_{01m}(\vec \rho)\varphi_{000}(\vec r)|{3\over 2},
t>|{1\over 2},\sigma>_+$\\
 & \\
 & $+\varphi_{000}(\vec \rho)\varphi_{01m}(\vec r)|{3\over 2},t>|{1\over 2}
,\sigma>_-\}$ \\
& \\ \hline
 & \\
$N(1650)$ & ${1\over \sqrt{2}}\sum_{m\sigma}(1,{3\over 2},m,\sigma|{1\over 
2},s)\{\varphi_{01m}(\vec \rho)\varphi_{000}(\vec r)|{1\over 2},t>_+$\\
 & \\
 & $+\varphi_{000}(\vec \rho)\varphi_{01m}(\vec r)|{1\over 2},t>_-\}|
{3\over 2},\sigma>$\\
 & \\ \hline
\end{tabular}
\end{center}

\newpage

\centerline{\bf Table 3}

\vspace{0.5cm}

Spin-flavor matrix elements of exchange current operators
for protons and neutrons with $s_z=+1/2$.

\vspace{1cm}

\begin{center}
\begin{tabular}{|c|c|c|} \hline
 & &\\
Operator & $<p,{1\over 2}|0|p,{1\over 2}>$ & $<n,{1\over 2}|0|
n,{1\over 2}>$\\
 & &\\ \hline
 & &\\
$(\vec \tau^1\cdot \vec \tau^2-\tau_3^1\tau_3^2)(\sigma_3^1+\sigma_3^2)$ &
4/3 & 4/3\\ 
 & &\\ \hline
 & & \\
$(\vec\tau^1\times \vec \tau^2)_3(\vec \sigma^1\times \vec \sigma^2)_3$
& -4 & 4\\ 
 & &\\ \hline
 & & \\
$\sigma^1_3+\sigma_3^2$ & 2 & 2\\
 & & \\ \hline
 & & \\
$\sigma_3^1\tau_3^1
+\sigma_3^2\tau_3^2$ & ${10\over 3}$ & $-{10\over 3}$\\
 & & \\ \hline
 & & \\
$\sigma^1_3\tau^2_3+\sigma^2\tau^1_3$ & $-{2\over 3}$ & ${10\over 3}$\\
 & & \\ \hline
 & & \\
$(\sigma^1_3+\sigma^2_3)\tau^1\cdot \tau^2$ & 2 & 2\\
 & & \\ \hline
\end{tabular}
\end{center}

\newpage

\centerline{\bf Table 4}

\vspace{0.5cm}

Mesonic contributions to the anapole moment of the proton
and the neutron.

\vspace{0.5cm}

\begin{center}
\begin{tabular}{|l|l|l|} \hline
 && \\
&proton&neutron\\
&&\\ \hline
&&\\
$\pi$ loops & $+3.22 \cdot 10^{-8}$& $+1.45\cdot 10^{-8}$\\
 && \\
$\pi$ exchange & $-1.15 \cdot 10^{-8}$&$-1.15\cdot 10^{-8}$\\
  && \\
$\pi$ polarization & $-0.19 \cdot 10^{-8}$&$-0.02\cdot 10^{-8}$\\
 && \\\hline
 && \\
$\rho$ loops & $-0.44 \cdot 10^{-8}$&$+1.52\cdot 10^{-8}$\\
 && \\
$\rho$ exchange & $-1.79 \cdot 10^{-8}$&$+0.73\cdot 10^{-8}$\\
 && \\
$\rho$ polarization&$+0.088\cdot 10^{-8}$&$-0.047\cdot 10^{-8}$\\ 
&&\\
 \hline
  && \\
$\omega$ loops & $-0.097 \cdot 10^{-8}$&$+0.34\cdot 10^{-8}$\\
 && \\ \hline
  && \\
$\rho \pi$ loops &$-0.67\cdot 10^{-8}$ &$-0.65\cdot 10^{-8}$\\
 && \\ \hline
  && \\
$\omega\pi$ loops &$+0.12\cdot 10^{-8}$&$-0.035\cdot 10^{-8}$ \\
& & \\ \hline
 & & \\
Total &$-0.90\cdot 10^{-8}$&$+0.68 \cdot 10^{-8}$ \\
& & \\ \hline
\end{tabular}
\end{center}

\newpage
\centerline{{\bf Figure Captions}}
\vspace{1cm}
Figure 1. Pion loop contributions to the anapole form factors
of the constituent quarks. The square vertex represents the parity 
violating pion-quark coupling (2.1a). To the current operator
should be added the contributions with the PV and PC hadronic
vertices permuted.

\vspace{1cm}

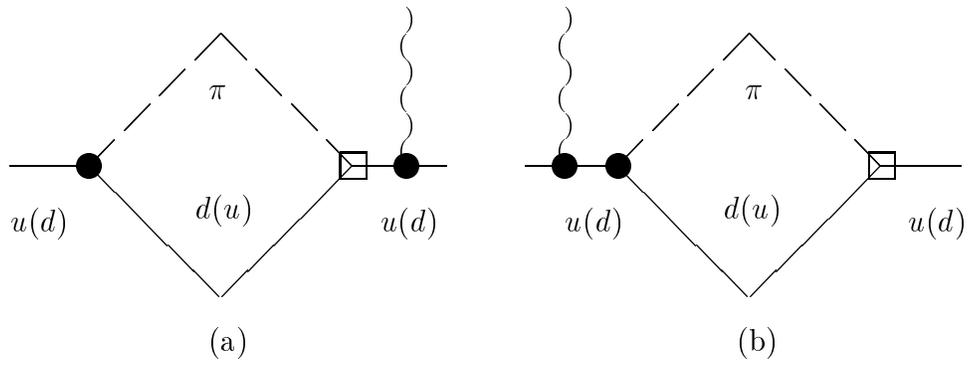
Figure 2. Self energy contributions to the anapole moments of the
constituent quarks that are required for current conservation.
The square vertex represents the parity 
violating pion-quark coupling (2.1a). 
To these should be added the corresponding contributions
with the PV and PC hadronic vertices permuted.

\vspace{1cm}

Figure 3. Calculated pion loop contributions to the anapole form 
factor of the proton and the neutron.

\vspace{1cm}

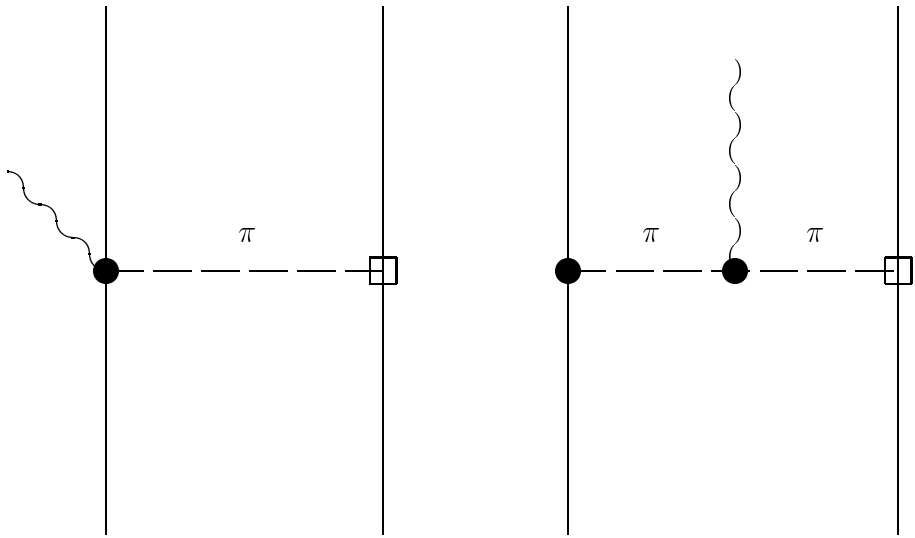
Figure 4. Pion exchange current contributions to the anapole
form factors of the nucleons. The fermion lines represent
constituent quarks. The square vertex represents the PV
pion-quark coupling (2.1a). The inclusion of the terms with
the PV and PC hadronic vertices should be understood.

\vspace{1cm}

Figure 5. Pion exchange current and polarization current
contributions to the anapole form factor of 
the proton. The PV pion exchange current contribution is denoted
"PI EXC" and the pion exchange induced polarization
current contribution is denoted PI POL. The curves A and B where 
obtained with the
values $\omega=$311 MeV and $\omega=$1240 MeV respectively
for the quark wave function parameter $\omega$ (3.5)

\vspace{1cm}

Figure 6. Pion exchange current and polarization current
contributions to the anapole form factor of 
the neutron. The PV pion exchange current contribution is denoted
"PI EXC" and the pion exchange induced polarization
current contribution is denoted PI POL. The curves A and B where 
obtained with the
values $\omega=$311 MeV and $\omega=$1240 MeV respectively
for the quark wave function parameter $\omega$ (3.5)

\vspace{1cm}

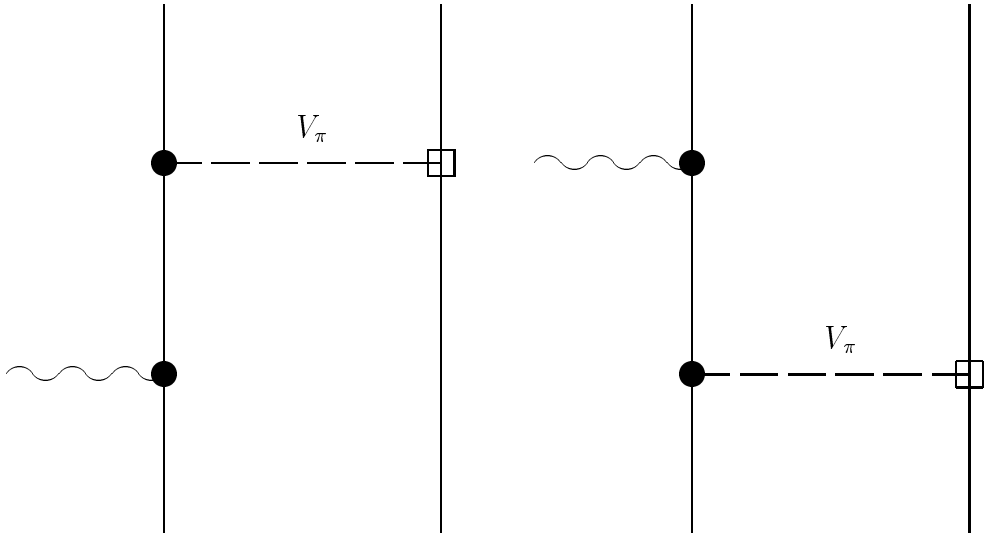
Figure 7. Schematic representation of the pion polarization current
that arises from the PV pion exchange interaction (3.3)
induced negative parity admixture in the nucleon wave
function.
The presence additional terms with the PC and PV pion-quark vertices
permuted should be understood.

\vspace{1cm}

Figure 8. Combined pion exchange current and polarization current
contributions to the anapole form factors of 
the proton and the neutron.

\vspace{1cm}

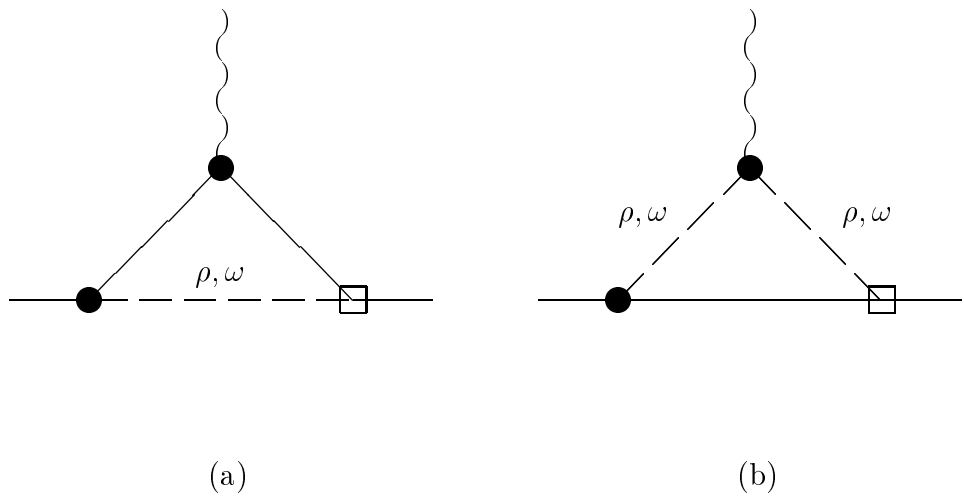
Figure 9. Vector meson loop contributions to the anapole
form factors of constituent quarks. The square vertex represents
the parity-violating couplings of vector mesons to constituent
quarks ((4.2) and (4.12). These should be combined with the
loop amplitudes with the PV and PC hadronic vertices permuted.

\vspace{1cm}

Figure 10. $\rho-$ and $\omega-$ meson loop 
contributions to the anapole 
form factors of the proton and the neutron. 

\vspace{1cm}

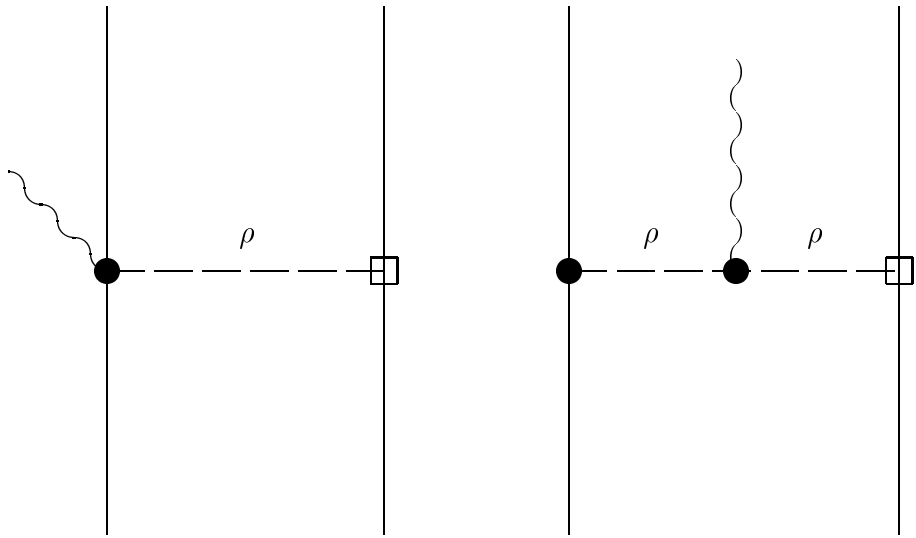
Figure 11. Parity violating $\rho$ meson exchange current 
contributions to the anapole form factors of the nucleon.
The square vertex represents the PV $\rho$ meson coupling
to constituent quarks (4.2). The inclusion of the terms with
the PV and PC hadronic vertices should be understood.

\vspace{1cm}

Figure 12. $\rho-$meson exchange current (EXC) and polarization
current contributions (POL) to the anapole form factor 
of the proton. The curves A and B where obtained with the
values $\omega=$1250 MeV and $\omega=$311 MeV respectively
for the quark wave function parameter $\omega$ (3.5).

\vspace{1cm}
Figure 13. $\rho-$meson exchange current (EXC) and polarization
current contributions (POL) to the anapole form factor 
of the neutron. The curves A and B where obtained with the
values $\omega=$1250 MeV and $\omega=$311 MeV respectively
for the quark wave function parameter $\omega$ (3.5).

\vspace{1cm}

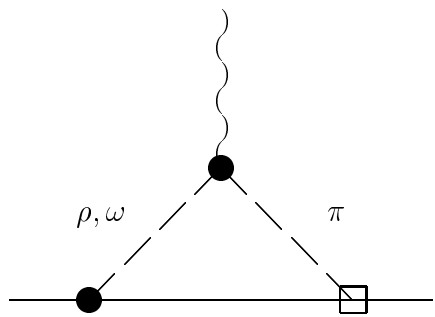
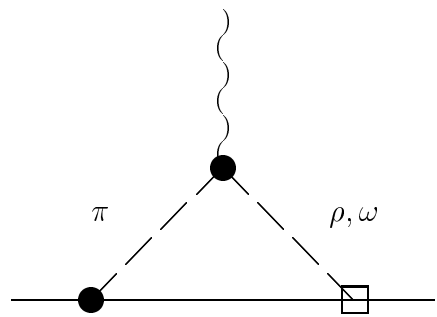
Figure 14. Parity violating $\rho\pi$
and $\omega\pi$ loop contributions to the
anapole form factors of the nucleons.
The square vertex represents the parity 
violating pion-quark coupling (2.1a). 

\vspace{1cm}

Figure 15. The $\rho\pi$ and $\omega\pi$ loop contributions
to the anapole form factors of the proton and the 
neutron. 

\vspace{1cm}

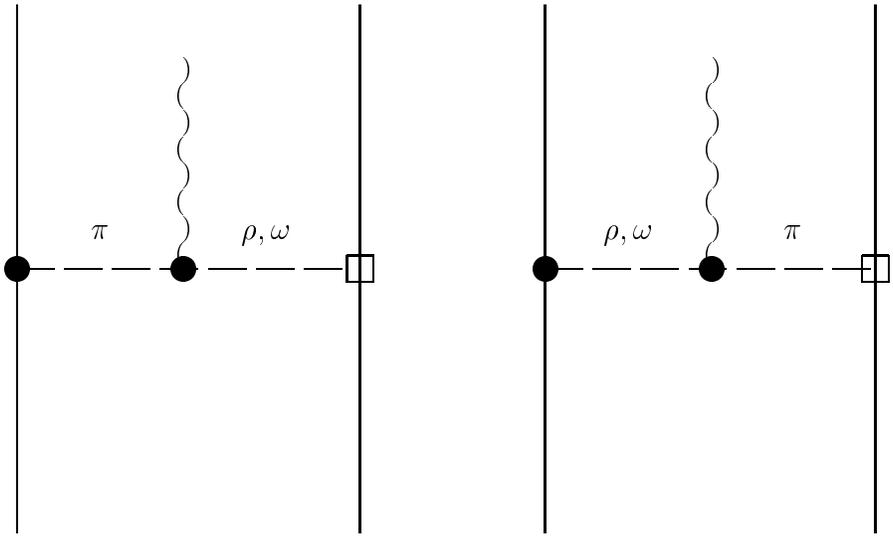
Figure 16. Mixed vector-pseudoscalar exchange current operators.

\vspace{1cm}

Figure 17. The net meson loop and exchange current contributions
to the anapole form factor of the proton and the neutron.
The results calculated without constituent quark form
factors are shown separately.

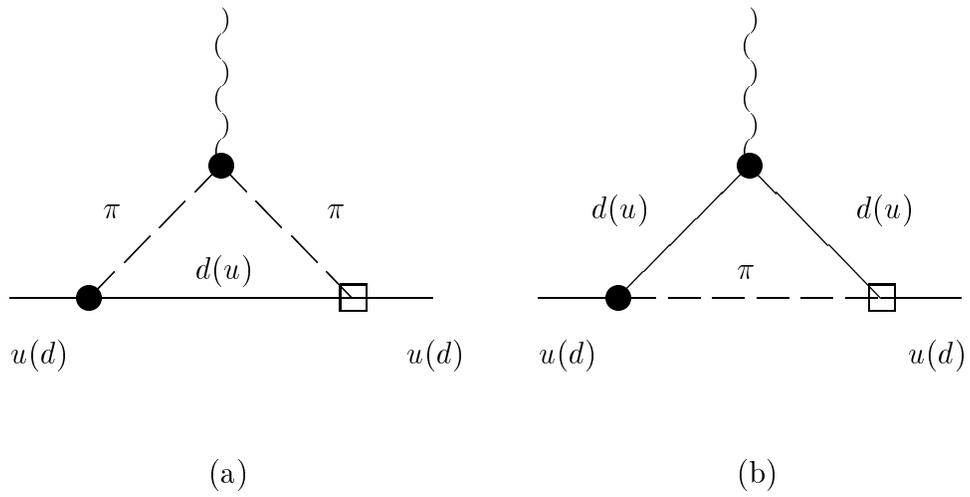
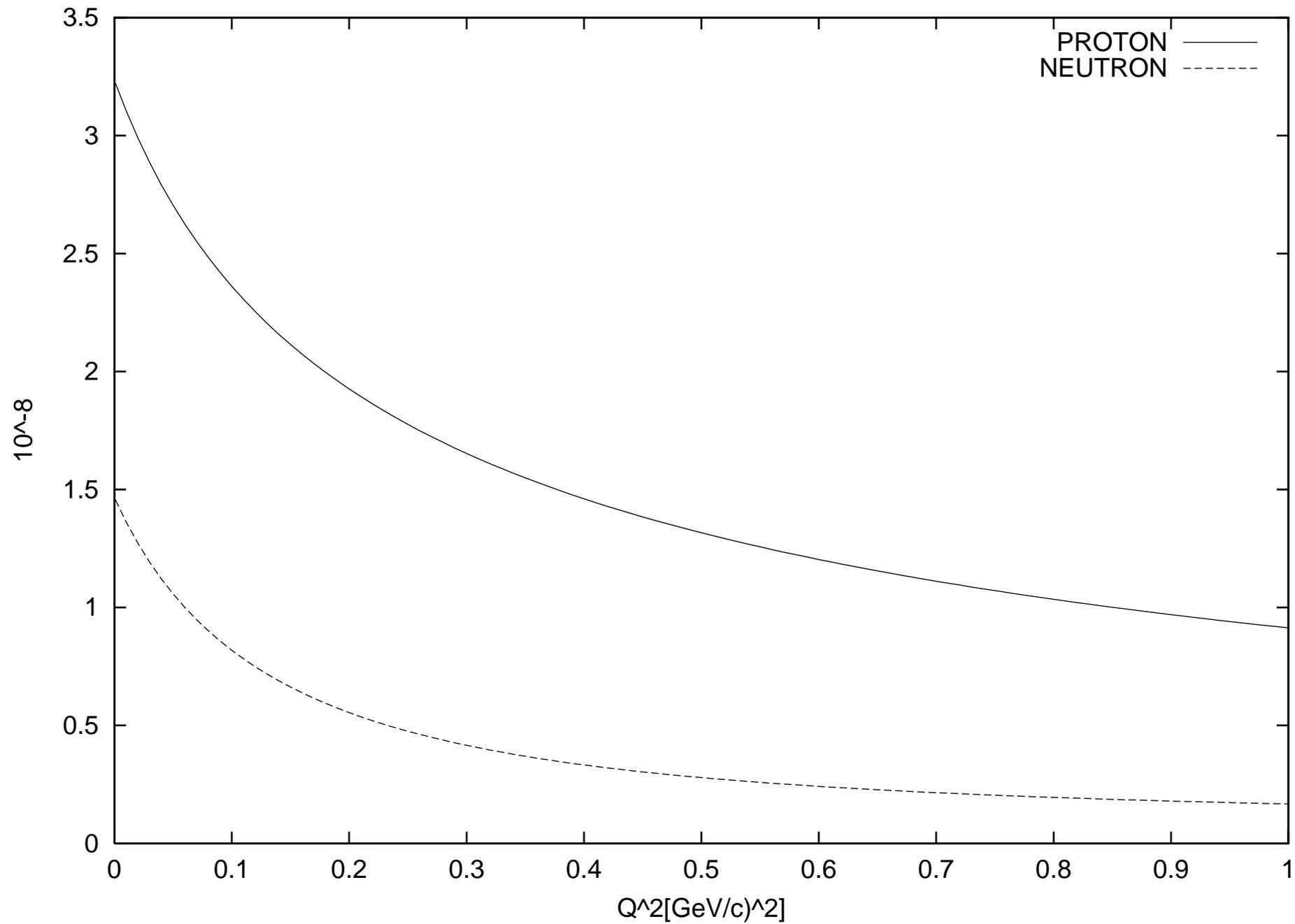
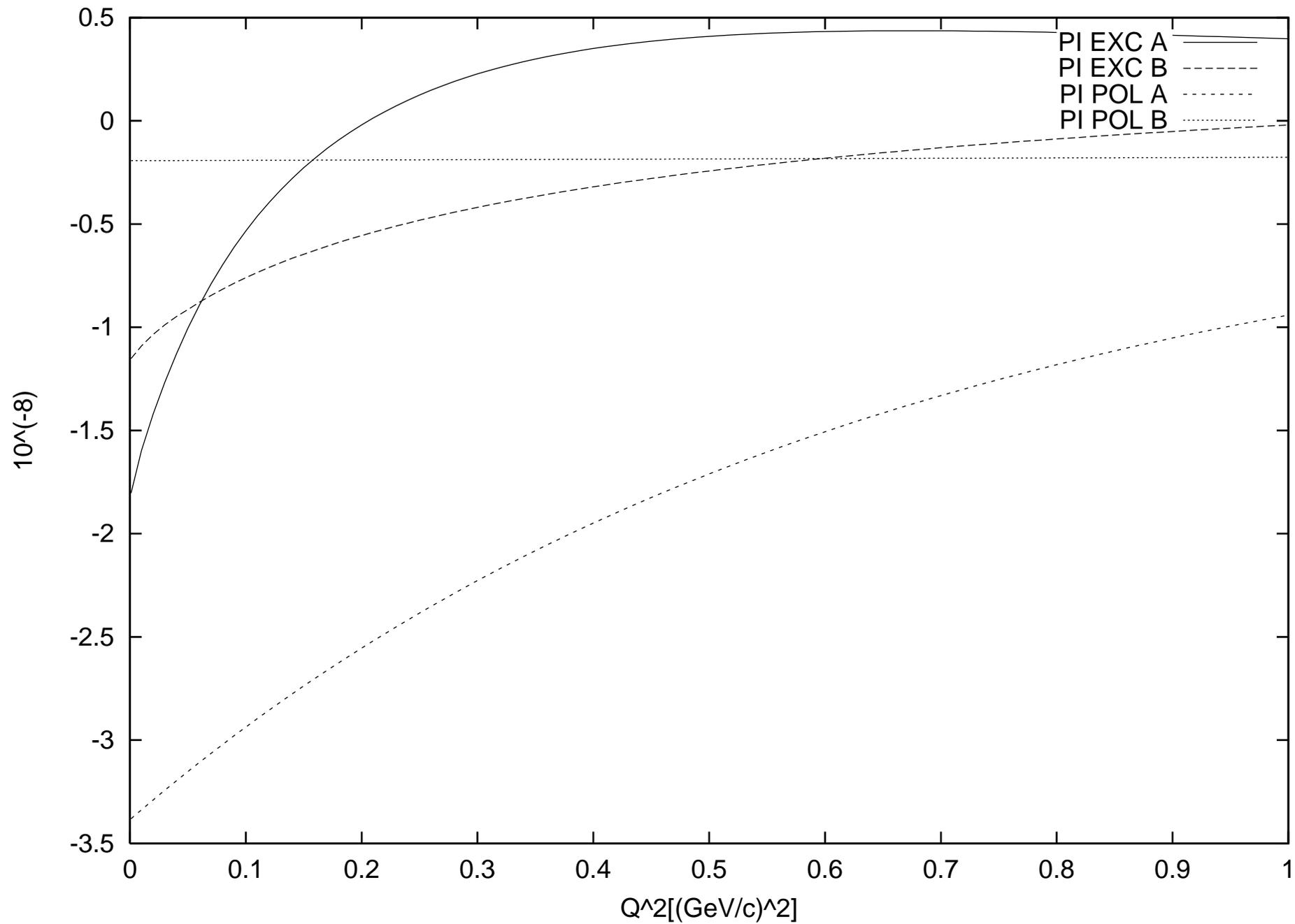
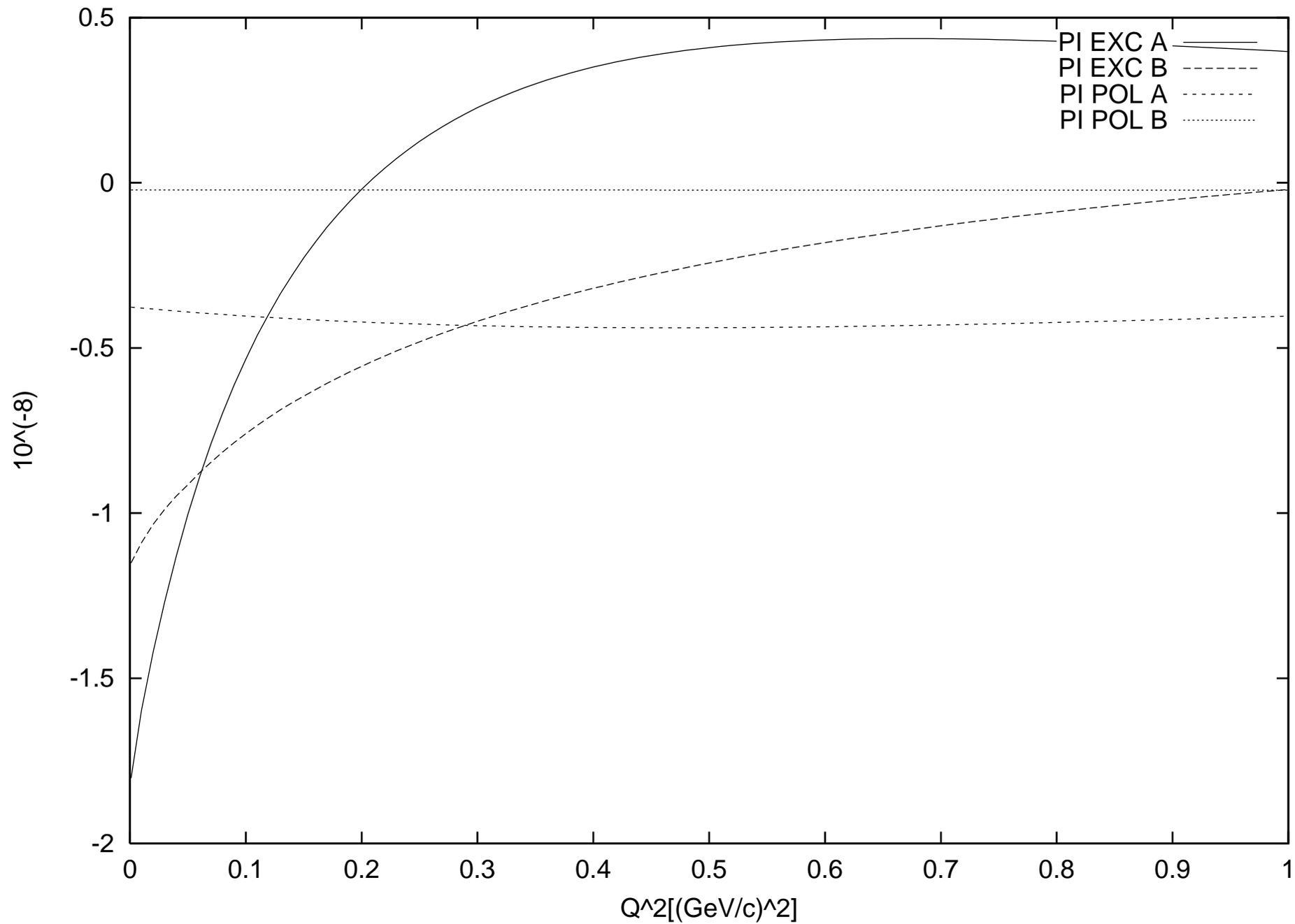
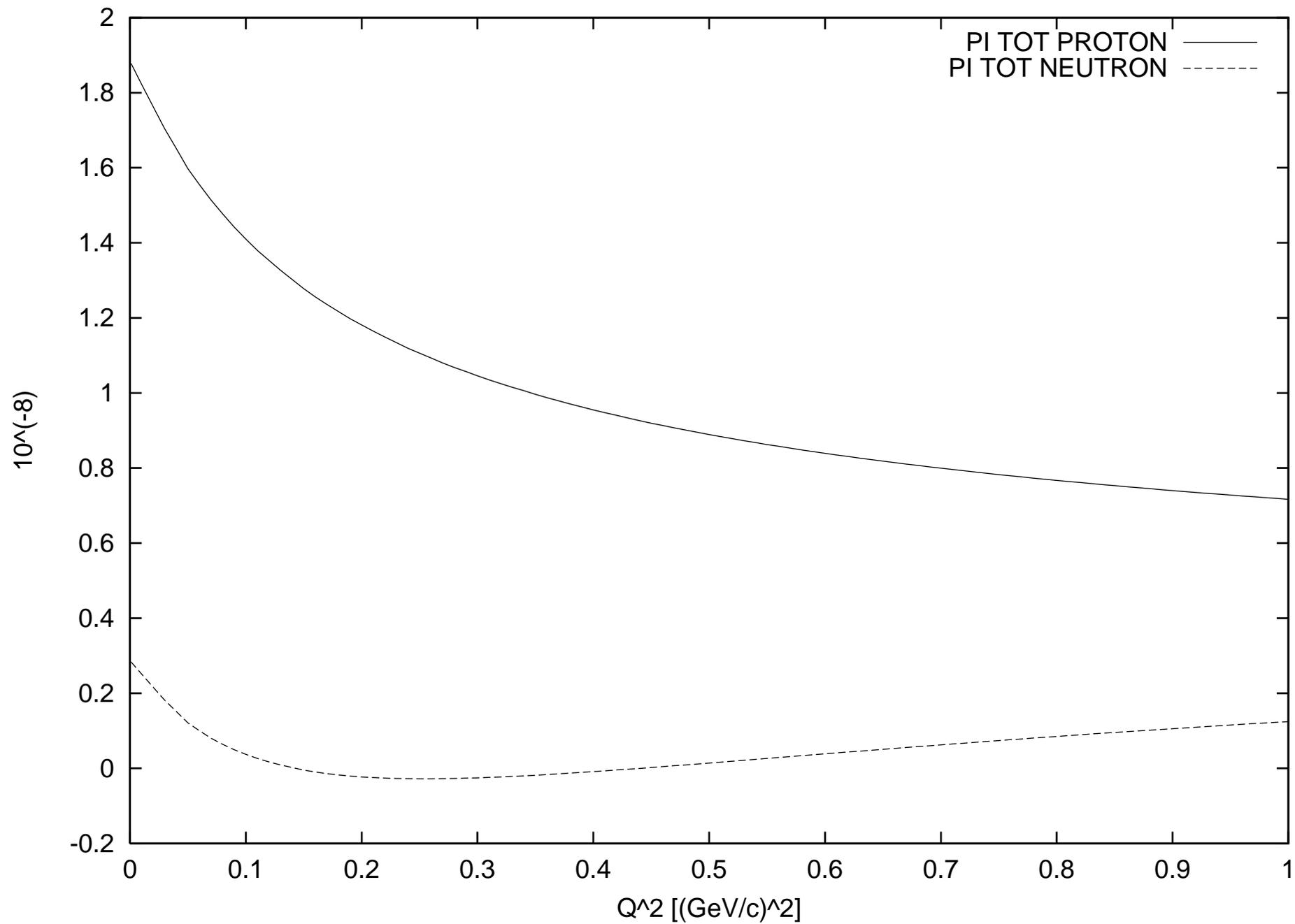
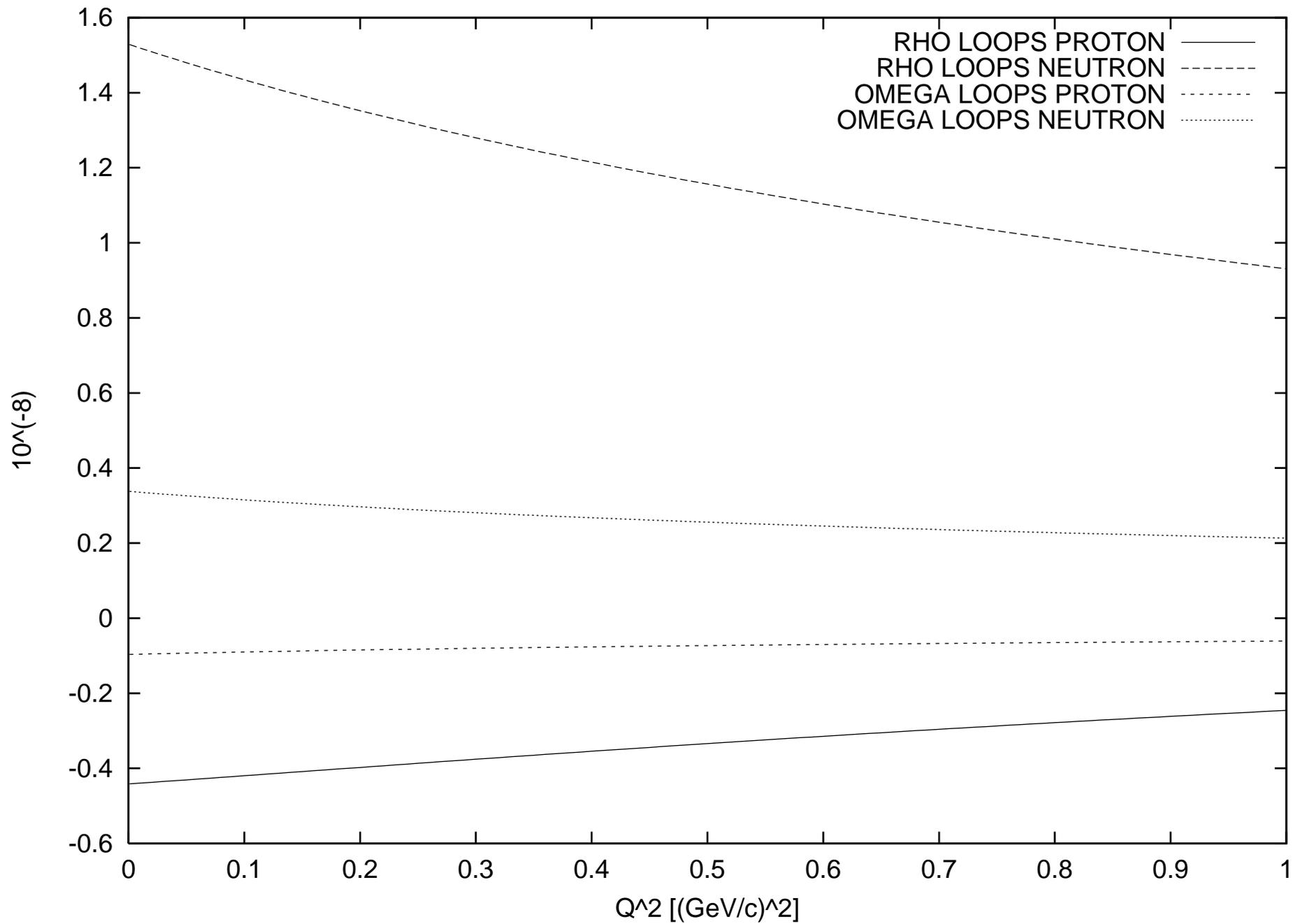
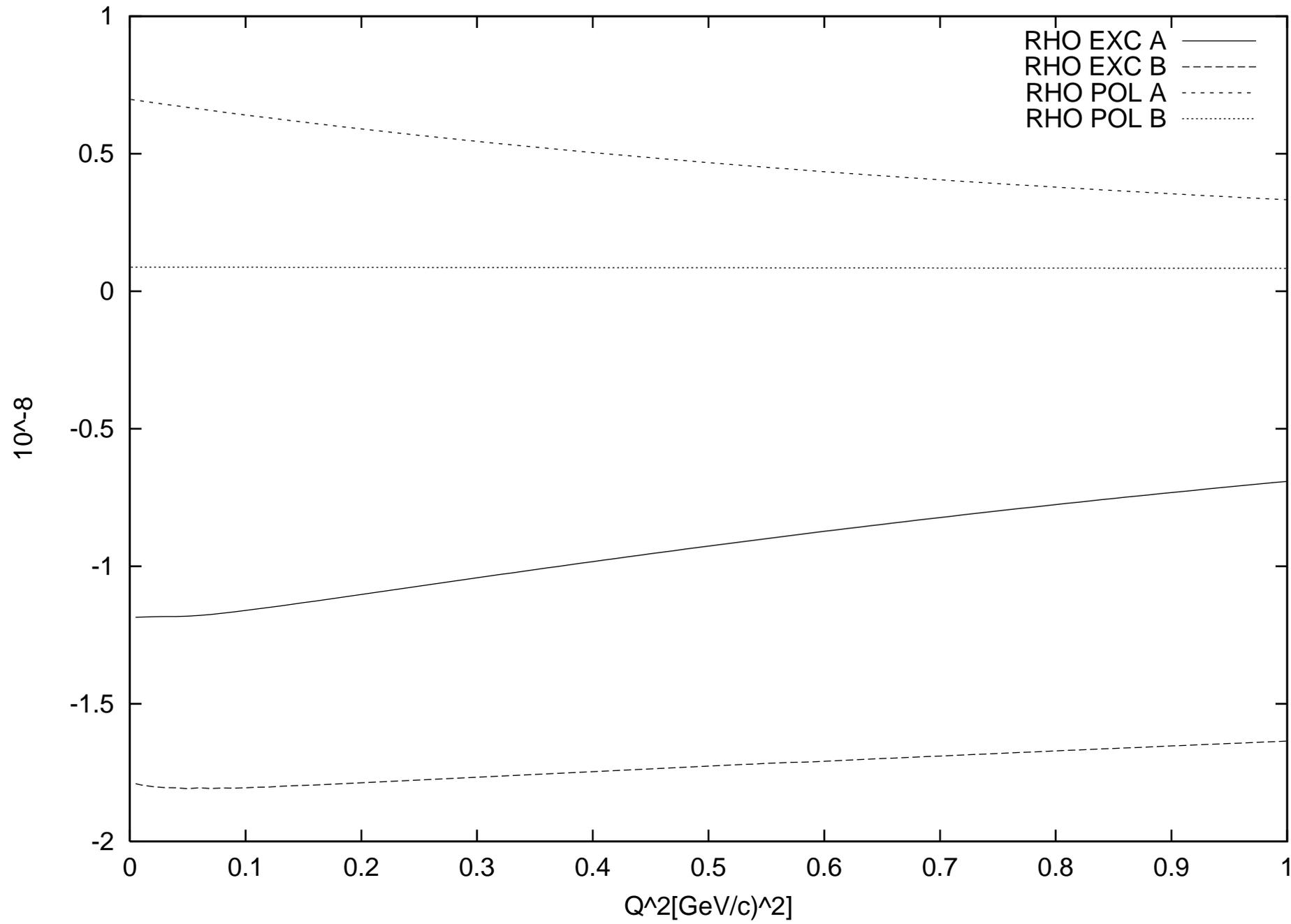
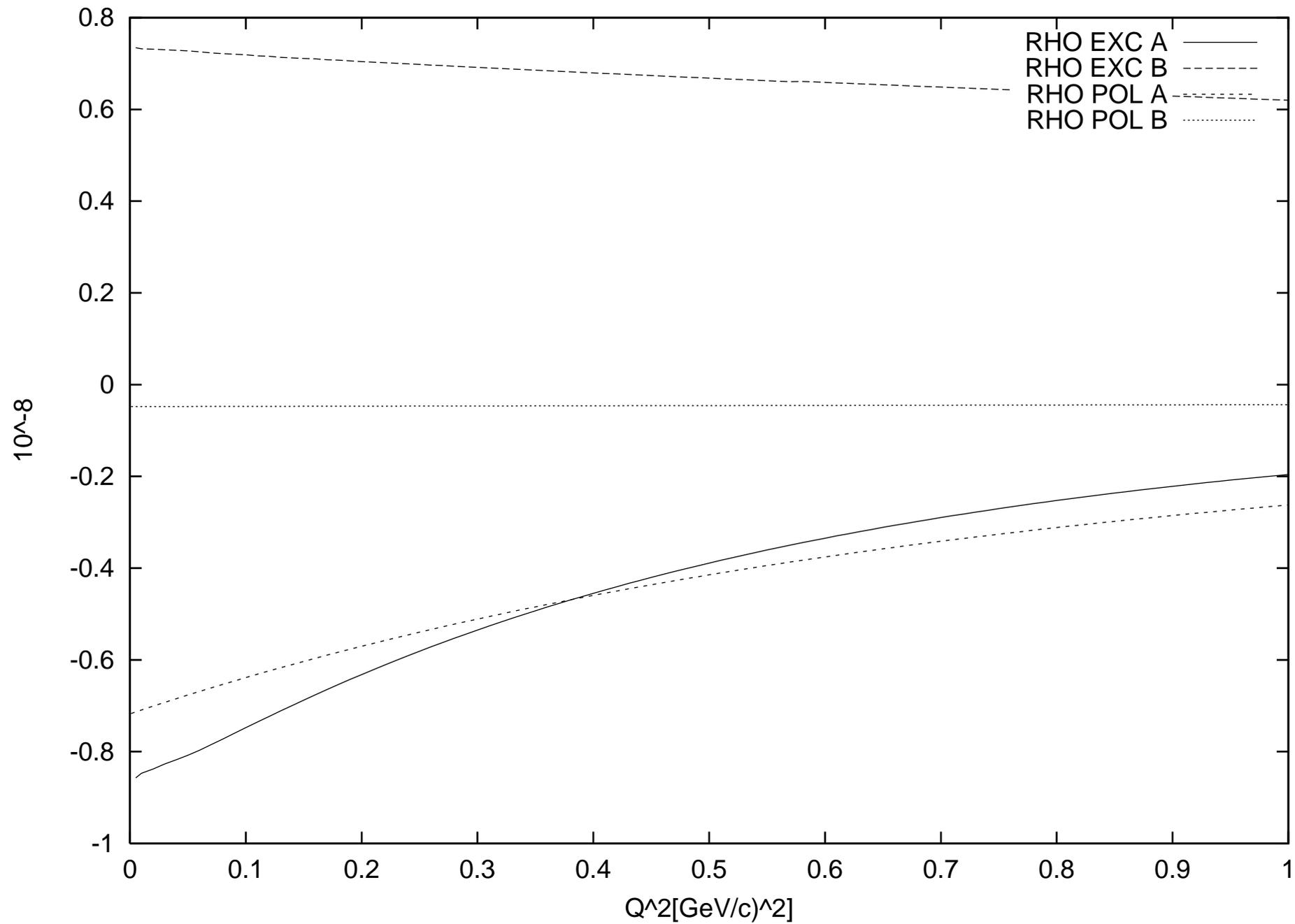
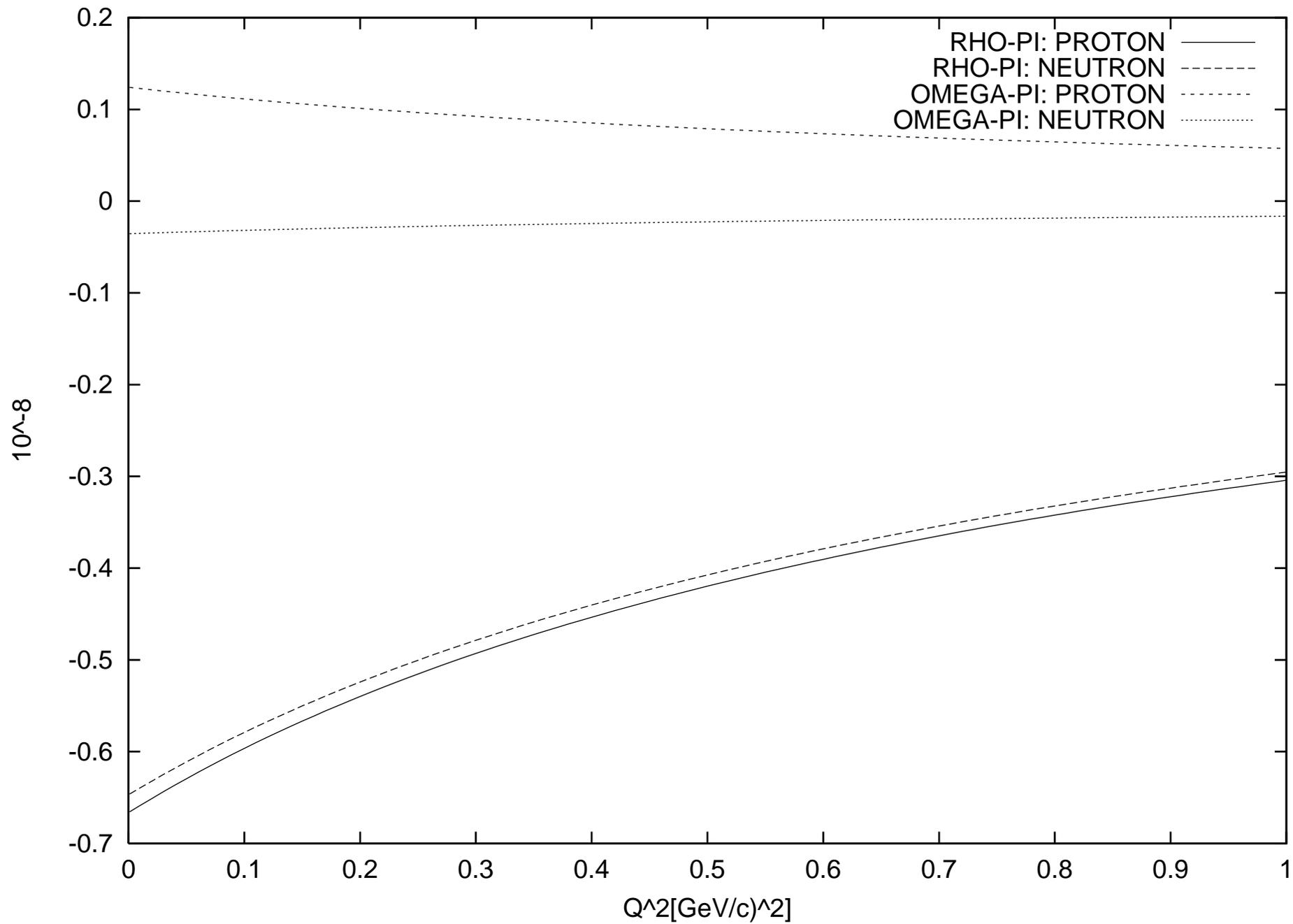
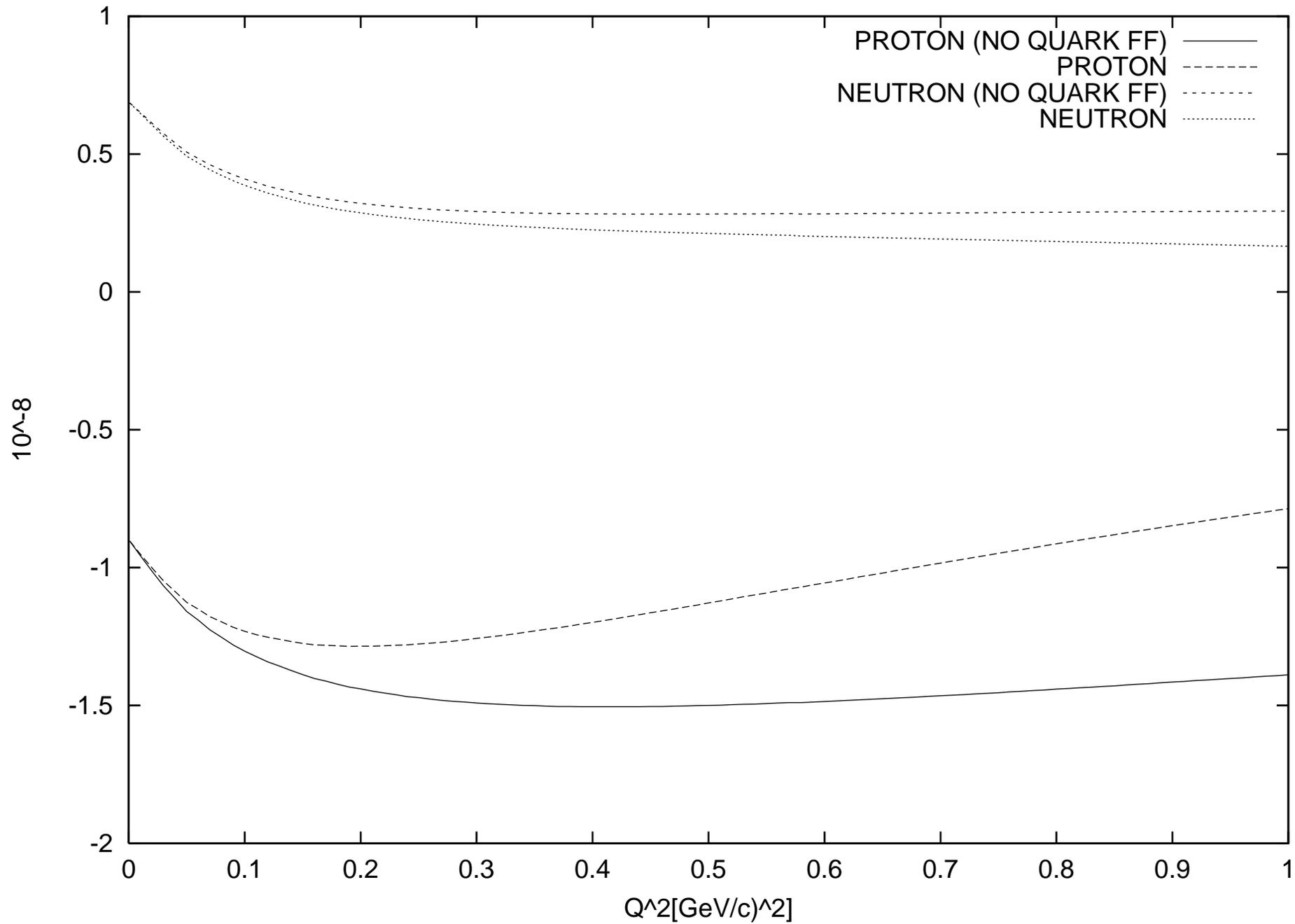
\end{document}